# Tracking the Photoinduced Dynamics of a Dark Excitonic State in Single-Layer $WS_2$ via Resonant Autler-Townes Splitting

Angela Montanaro[1,2,3,*], Francesco Valiera[4,*], Francesca Giusti[1,2], Francesca Fassioli[3], Chiara Trovatello[5,6], Giacomo Jarc[1,2,3], Enrico Maria Rigoni[1,2,3], Fang Liu[7,8], Xiaoyang Zhu[7], Stefano Dal Conte[5], Giulio Cerullo[5,9], Martin Eckstein[4,10], and Daniele Fausti[1,2,3,†]

*1. Department of Physics, Università degli Studi di Trieste, Trieste, Italy.*
*2. Elettra Sincrotrone Trieste, Trieste, Italy.*
*3. Department of Physics, University of Erlangen-Nürnberg, Erlangen, Germany*
*4. Institute of Theoretical Physics, University of Hamburg, Hamburg, Germany*
*5. Dipartimento di Fisica, Politecnico di Milano, Milano, Italy*
*6. Department of Mechanical Engineering, Columbia University, New York, 10027, NY, USA*
*7. Department of Chemistry, Columbia University, New York, 10027, NY, USA*
*8. Department of Chemistry, Stanford University, Stanford, CA 94305, USA*
*9. Istituto di Fotonica e Nanotecnologie, Consiglio Nazionale delle Ricerche, Milano, Italy*
*10. The Hamburg Centre for Ultrafast Imaging, Hamburg, Germany*

[*] These authors contributed equally to the work.
[†] Correspondence to: daniele.fausti@fau.de

**ABSTRACT**
Excitons in a monolayer transition metal dichalcogenide (1L-TMD) are highly bound states characterized by a Rydberg-like spectrum of discrete energy levels. Among these, states with odd-parity are known as dark excitons due to selection rules, which make their stationary and transient characterization challenging using linear optical techniques. Here, we demonstrate that the dynamics of a $2p$ dark excitonic state in 1L-$WS_2$ can be directly retrieved by measuring the Autler-Townes splitting of bright states in a three-pulse experiment. The splitting of the bright $1s$ excitonic state, observed by detuning a mid-infrared control field across the $1s$–$2p$ transition, provides an accurate characterization of the $2p$ state. Following carrier photoinjection, we observe a qualitatively different dynamics of the $1s$ and $2p$ levels, which is indicative of symmetry-dependent screening and exciton-exciton interactions. These findings provide new insights into many-body effects in TMDs, offering potential avenues for advancing the next generation optoelectronics.

Monolayer transition metal dichalcogenides (1L-TMDs) have gained attention due to their unique electronic and optical properties [1-3]. Their strong nonlinear response [4,5] and excitonic photophysics [6,7] make them ideal for optoelectronic devices that exploit strong light-matter interactions. In these atomically thin semiconductors, efficient exciton-light coupling arises from reduced dimensionality and weak dielectric screening, resulting in high exciton binding energies (2). This strong Coulomb interaction yields a Rydberg-like spectrum of discrete excitonic levels, reminiscent of hydrogen atoms but influenced by the material's environment [8].
Excitonic states are categorized as "bright" or "dark" based on their interaction with light. Bright states (even parity) interact strongly with photons and appear in linear optical experiments, while dark states (odd parity) respond weakly, making them difficult to characterize [9-11].
Recently, advanced spectroscopic tools and/or external perturbations (e.g., magnetic fields, strain, nonlinear probes) have been used to bypass selection rules and reveal dark states in different families of 1L-TMDs [12-17]. In particular, two-photon spectroscopy has uncovered the odd-parity $2p$ and $3p$ dark states of the A exciton in 1L-$WS_2$ [12], which have been found robust against temperature and dielectric variations, suggesting that static perturbations do not alter the $p$-symmetry states.
However, this robustness may not apply to impulsive perturbations. Studies show that early exciton dynamics after photoexcitation is dominated by phase space filling, bandgap renormalization, and many-body interactions [7,18-22]. The simultaneous occurrence of these effects, each producing distinct and often opposing optical signatures, complicates the task of disentangling their respective contributions.

So far, experimental investigations have mainly focused on bright excitons, where bandgap shrinkage and reduced oscillator strength have been found to dominate the renormalization [18,22,23]. However, the ultrafast dynamics of dark excitons remains largely unexplored. In particular, the spatially dependent screening from photoexcited carriers and resulting exciton-exciton interactions may impact wavefunctions differently depending on their symmetry.

Here, we compare the dynamics of the $1s$ and $2p$ states of the A exciton in 1L-$WS_2$ following an impulsive above-gap photoexcitation at 3.1 eV ($\omega_{pump}$[1] in Fig. 1a). To probe the $2p$ state, we developed a three-pulse technique (see [24] Note 1 and Ref. [25]) that leverages the strong oscillator strength of the bright $1s$ state to induce an Autler-Townes splitting in the optically active transition. This effect arises in the presence of a mid-infrared (mid-IR) control field ($\omega_C$ in Fig. 1a) resonant with the $1s$–$2p$ transition. The transient transmissivity of the sample is then probed by a broadband visible pulse ($\omega_p$ in Fig. 1a), whose spectrum covers the bright $1s$ transition, which is dispersed and frequency-resolved after transmission. The large area 1L-$WS_2$ sample was obtained by gold-tape assisted exfoliation (see [24] Note 2 and Ref. [26]) and all the measurements have been carried out at room temperature.

The rationale of our approach is illustrated in Fig. 1b. The sample can be modelled as an ensemble of independent 3-level ladder systems, as depicted in the leftmost panel. The allowed optical transition between the maximum of the valence band (VB) at the K point in the Brillouin zone ($|1\rangle$) and the $1s$ exciton state ($|2\rangle$) occurs at energy $\omega_{12}$ = 2.029 eV [27], while we indicate with $\omega_{23}$ the energy separation between the $2p$ ($|3\rangle$) and the $1s$ levels, which is expected to lie in the mid-IR range [12]. The transition between $|3\rangle$ and $|1\rangle$ is instead dipole-forbidden, thus making $|3\rangle$ a dark state. The absorption by the bright 1s exciton is visible in linear optical transmission measurements (green line in leftmost panel of Fig.1c).
When an intense pulsed control field $\omega_C$ is applied and is tuned on resonance with the transition $|2\rangle \rightarrow |3\rangle$, a splitting of the transition into two “dressed” states occurs. This effect – which is the AC counterpart of the static Stark effect – is referred to as Autler-Townes splitting (ATS) [28] and arises from classical strong light-matter coupling. In the two-pulse control scheme (Fig. 1b, middle panel), $\omega_C$ acts as a gate field for ATS to occur and $\omega_p$ probes the transient transmissivity of the system. The differential transmissivity ($\Delta T/T$) map in the middle panel of Fig. 1c displays an increased transmissivity in correspondence with level $|2\rangle$, confirming a suppressed absorption of the $1s$ excitonic state and the formation of doublets, appearing as blue regions of decreased transmissivity at both lower and higher energies with respect to $\omega_{12}$. This effect can be better visualized by plotting the $\Delta T/T$ spectrum (grey line in the leftmost panel of Fig. 1c), where the horizontal dotted line indicates the $1s$ absorption peak at equilibrium.
Since the amplitude, the intensity and the central frequency of the ATS intrinsically depend on the spectral properties of level $|3\rangle$ (i.e., its energy $\omega_{23}$ and its coherence decay rate $\gamma_{13}$), optical measurements of the ATS spectrum in our sample provide indirect insights on the dark $2p$ state. This approach has previously been employed to resolve exciton fine structures in the K and K′ valleys in $MoSe_2$ by coupling the $1s$ and the $2p^+$ and $2p^-$ dark states using circularly polarized mid-IR pump pulses [29,30]. In this work, we measure the ATS spectrum at varying time delays between the visible pump pulse and the mid-IR control field (three-pulse control in Fig. 1b). This allows us to achieve time-resolution and reconstruct the pump-induced dynamics of the $2p$ state following the above-gap photoexcitation. An example of a three-pulse $\Delta T/T$ map is plotted in the rightmost panel of Fig. 1c, where the above-gap photoexcitation is followed after 0.5 ps by the control field $\omega_C$. The ATS spectrum (black line in the leftmost panel of Fig. 1c) gets modified by the presence of free carriers injected by $\omega_{pump}$.

In order to formalize the experiment within an ATS framework, we developed a semi-classical model which couples a 3-level ladder system with both the mid-IR and the probe field (see [24] Note 3).The interaction Hamiltonian of the model is given by the Rabi interaction for each of the two allowed transitions:

---

[1] We will always assume in the following that $\hbar = 1$.

$$H_{Rabi}(t) = -\frac{\Omega_p}{2}\left(e^{-i(\omega_{12}+\Delta_p)t}|2\rangle\langle 1| + h.c.\right) - \frac{\Omega_C}{2}\left(e^{-i(\omega_{23}+\Delta_C)t}|3\rangle\langle 2| + h.c.\right)$$

[Eq. 1]

Where we have denoted by $\Delta_p = \omega_p - \omega_{12}$ and $\Delta_C = \omega_C - \omega_{23}$ the detunings of the probe and the control fields with respect to the excitonic transitions. We have also denoted by $\Omega_p = \mu_{12}E_p$ and $\Omega_C = \mu_{23}E_C$ the Rabi frequencies of the probe ($E_p$) and the mid-IR ($E_C$) fields, where $\mu_{ij}$ is the transition dipole moment of the $|i\rangle \rightarrow |j\rangle$ transition.

By computing the coherent evolution of the system within the density matrix formalism, we obtained an expression of the complex susceptibility for the transition from the ground state to the 1$s$ level, namely the transition monitored in our experiment as a function of the probe photon energy (Eq. S3).

It should be stressed that the framework we are discussing here is analogous to the one established in the context of light-driven atomic or molecular systems, in which the coherent preparation by a laser excitation leads to the well-known phenomenon of electromagnetically-induced transparency (EIT) [31,32]. In our case, however, the intrinsic inhomogeneous broadening of the excitonic transitions and their fast dephasing times do not satisfy the conditions for EIT to occur, specifically $\gamma_{13} < \gamma_{12}$, where $\gamma$ denotes the coherence decay rate of the states. Notably, our measurements reveal that the dephasing rate of the 2$p$ state is approximately three times larger than the one of the 1$s$ exciton, as it will be demonstrated below. This indicates that a coherent superposition cannot be sustained in the system under study, and destructive quantum interference does not occur. A study of the parameter space of the model is given in Ref. [24] (Note 3).

Fig. 2a shows the $\Delta T/T$ spectra measured at the temporal overlap between $\omega_C$ and $\omega_p$ by varying the mid-IR photon energy, while keeping the excitation fluence fixed. The emergence of a resonance is evident: the amplitude of the transparency window reaches its maximum at $\omega_C = 155$ meV (light green curve) and decreases for both higher and lower mid-IR pump photon energies, as shown in the inset of Fig. 2a. This reduction in amplitude is accompanied by a gradual shift of the transparency window toward higher probe photon energies as $\omega_C$ decreases (from the purple to the red curve). The trend observed is consistent with our simulations with a varying detuning of the mid-IR field (Fig. S1d in [24]).

To quantitatively describe this behavior, a global fit of the curves in Fig. 2a was performed using Eq. S3. The fit function includes three free parameters: $\Omega_C$, $\Delta_C$, and $\gamma_{13}$, with the constraint that $\gamma_{13}$ - being an intrinsic material property - remains constant across all curves. The parameters related to the transition from the 1$s$ exciton to the ground state ($C$, $\omega_{12}$ and $\gamma_{12}$) are instead constrained by fitting the static absorption (see [24] Note 4, [33]) and kept constant (see [24] Note 5 for details on the global fitting procedure). From the fit, we estimate $\gamma_{13} = 65$ meV, corresponding to a dephasing rate of the 2$p$ state that is approximately three times larger than that of the 1$s$ state.

Fig. 2b and 2c present the extracted Rabi frequency and control field detuning, respectively, as functions of the incident mid-IR photon energy. In Fig. 2d we plot three representative spectra of the mid-IR pulse.

The Rabi frequency (Fig. 2b) exhibits a maximum at $\omega_C = 155$ meV, indicating that the coupling between the mid-IR control field and the $|2\rangle \rightarrow |3\rangle$ transition is maximized under this condition. Similarly, the fitted detuning at this pump energy is $\Delta_C = 0$ (Fig. 2c), implying that the control field is resonant with the 1$s$–2$p$ energy separation at a driving photon energy of 155 meV. This provides a direct measurement of the intraexcitonic 1$s$–2$p$ transition energy. We emphasize that the fitted value of $\Delta_C$ is consistent with the expected detuning, given $\omega_{23} = 155$ meV (indicated by the shaded grey line in Fig. 2c). These independent measurements of the detuning serves as a critical validation of the model and the fitting procedure.

Furthermore, by investigating the transiently induced ATS under resonant mid-IR driving ($\omega_C = 155$ meV) across various fluences (see [24] Note 5), we extracted the Rabi frequency versus $\sqrt{\phi_C}$ plotted in Fig. 2e. The curve initially follows a linear trend (grey line) due to its proportionality to the field amplitude. At higher fluences, $\Omega_C$ saturates—also reflected in the ATS spectral amplitude (Fig. S7)—likely due to Pauli blocking from resonant driving. Notably, this saturation is absent when the mid-IR field is detuned from the $1s$–$2p$ resonance (Fig. S9), where a linear dependence persists even at high fluences.
Our estimated value for the $1s$-$2p$ energy separation is approximately 80 meV smaller than that reported in Ref. [12], which was determined using two-photon excitation spectroscopy. We reckon that this discrepancy may originate from differences in the experimental methodologies employed. In Ref. [12], the authors derived the energy separation by measuring photoluminescence as a function of the two-photon energy required to excite the $2p$ state. In contrast, our approach involves direct measurement of the absorption between the $1s$ and $2p$ energy levels. Notably, the $1s$-$2p$ energy separation obtained in our study coincides with the onset of the $2p$ photoluminescence peak observed in the two-photon excitation spectroscopy measurements reported in Ref. [12].

The characterization of the mid-IR-induced ATS as a function of $\omega_C$ photon energy and fluence provides a validation of our model and enables an accurate estimate of the $1s$-$2p$ energy separation. Thus far, our analysis assumes an initial equilibrium state before interaction with $\omega_C$, using static parameters of the $1s$ transition from equilibrium absorption. However, studying the photoinduced dynamics of the $2p$ state requires understanding the $1s$ exciton's response to the above-gap photoexcitation. Since the $1s$ energy and linewidth dynamically evolve under above-gap pumping, a full characterization is necessary to benchmark ATS at each time delay. As our model predicts only relative shifts between the $1s$ and $2p$ levels, the precise knowledge of the absolute $1s$ transition energy is crucial for quantifying the $2p$ renormalization. By measuring the transient transmissivity induced by a single pulse at $\omega_{pump} = 3.1\ eV$ and performing a differential fit of the data (see [24] Note 7), we extracted the temporal evolution of the parameters related to the $1s$ exciton. Upon photoexcitation, the $1s$ peak undergoes two major modifications: a redshift of its central energy by ~5 meV and a broadening of its lineshape (Fig. S10b and S10c). Both these effects occur in a time window shorter than 1 ps. A longer dynamics – associated to interactions with the lattice [34] - persists up to at least $t = 25$ ps (Fig. S13). The single pump results are consistent with previous measurements in the same compound and similar 1L-TMDs [18,20,22].

We used the temporal evolution of $\omega_{12}$ and $\gamma_{12}$ (Fig. S10) to benchmark the three-pulse experiment (see [24] Note 9 for details on the three-pulse differential acquisition, [25]). The inset of Fig. 3a shows three representative maps and corresponding ATS spectra with (black) and without (grey) above-gap photoexcitation. The panels refer to different $\Delta t$: for $\Delta t < 0$ the mid-IR precedes the visible pump, while for $\Delta t > 0$, it follows it. The $\Delta T/T$ spectra are taken at the temporal overlap between the probe and the mid-IR pulse. The data indicate that the visible photoexcitation, which exerts its strongest influence at the temporal overlap between the two pumps ($\Delta t = 0$, central panel), induces a quenching and a broadening of the ATS transparency window.
We scanned $\Delta t$ densely and, for each delay, the ATS spectra were fitted using time-dependent $1s$ parameters. The fitting shows that $\Delta_C$ is most affected, indicating a dynamic modulation of the $1s$–$2p$ energy gap. Using the temporal evolution of $\omega_{12}$ extracted from the experiments with single above-gap pump (Fig. S10b), we isolated the $2p$-level dynamics. Fig. 3a compares the pump-induced shifts of the $1s$ [$\delta_{1s}(\Delta t)$, purple] and $2p$ [$\delta_{2p}(\Delta t)$, pink] levels. We observe that also the $2p$ state experiences a pump-induced redshift, whose amplitude at early times is more than twice the one of the $1s$ state. Furthermore, the dynamics of $\delta_{2p}(\Delta t)$ is faster and, within 500 fs, the $2p$ level blueshifts by about 5 meV.

We propose a tentative explanation for this observation by considering the influence of spatially dependent ultrafast screening on wavefunctions with different symmetry and principal quantum number [35]. Figure 3b summarizes the temporal evolution of screening and exciton-exciton interactions following photoexcitation. At early delays, above-gap excitation injects free carriers, initially distributed homogeneously, enhancing effective screening—especially for excitonic states with larger electron-hole separation. Thus, higher-n states experience stronger renormalization, explaining the larger energy shift of the $2p$ state compared to $1s$ (Fig. 3a).
At later times, free carriers bind into excitons. The excitation density estimated ($1.2 \times 10^{13}$ cm$^{-2}$) suggests that screening is no longer dominated by free charges but also by exciton-exciton interactions.

As excitons localize, the in-plane screening becomes spatially inhomogeneous, affecting the $1s$ and $2p$ wavefunctions differently.
Conventionally, exciton-exciton interactions in 1L-TMDs are described within a semiclassical framework, treating excitons as neutral particles with finite polarizability that experience long-range dipole-dipole repulsion [36, 37], which results in a net blueshift of the excitonic energies. Recent theoretical work [37] further demonstrates that a blueshift can also arise from the fermionic nature of the exciton substructure, analogous to Pauli blocking of the fermionic phase space. This effect leads to a "hardcore" exciton-exciton repulsion, contributing to an increase in excitonic energies.
The observed $2p$ blueshift suggests that one or both these repulsive interactions become significant as carriers form excitons. The $1s$ state lacks a similar shift, likely due to its more localized wavefunction. The extended $2p$ exciton is more more susceptible to phase-space filling and repulsion in dense exciton regimes.

In conclusion, we introduced a novel methodology to probe the dynamics of dark states in 1L-TMDs by measuring the ATS of the bright exciton state. Comparing the dark exciton state dynamics with that of the bright one allowed us to identify the dominant screening mechanisms and isolate exciton-exciton interactions in the dark state renormalization. Notably, while we focus here on $p$-symmetry states in 1L-TMDs, the approach we developed could be generalized to the dynamics of any dark state dipole-connected to optically-bright states. Finally, we stress that accessing the dynamics of dark states in low dimensional systems offers a new fundamental insight on their relaxation pathways, with profound implications also for optoelectronics applications.

## CAPTIONS

***Figure 1: Time-dependent Autler-Townes splitting (ATS) to probe the dynamics of the 2p dark state.*** *a) Sketch of the experimental three-pulse setup. b) Conceptualization of the experiment, in which the sample is modelled as a 3-level system. We denoted by $\omega_{12}$ and $\omega_{23}$ the energies of the bright 1s and dark 2p excitons, respectively. The broadband probe ($\omega_p$, green arrow) is centered on the 1s state. In the two-pulse control scheme, the excitation by a mid-infrared control field ($\omega_C$, red arrow) resonant to the 1s-2p transition triggers an ATS of the bright 1s exciton, which is revealed by $\omega_p$, as shown in the transient ΔT/T map below. In the three-pulse protocol, a photoexcitation by an above-gap pump pulse ($\omega_{pump} = 3.1$ eV, blue arrow) impulsively renormalizes the excitonic spectrum. The ATS as a function of the time separation (Δt) between $\omega_C$ and $\omega_{pump}$ provides information on the pump-induced dynamics of the 2p dark state. c) Static transmission of 1L-$WS_2$ (empty circles) and related fit (green line). The black and the grey curves are the ATS spectra at the overlap between $\omega_C$ and $\omega_p$ (as indicated by the vertical lines on the color-coded maps) with and without the excitation by $\omega_{pump}$, respectively.*

***Figure 2: Resonant ATS in $WS_2$.*** *a) Spectra of the ATS at the overlap between $\omega_C$ and $\omega_p$ for different photon energies of the pump and fixed fluence $\phi_C = 50\ \mu J cm^{-2}$. The inset shows the maximum $\Delta T/T$ as a function of the mid-IR photon energy. b) Rabi frequency of the mid-IR control field as a function of its photon energy, as extrapolated from the fits of the curves in a). c) Extracted detuning of the mid-IR control field from the 1s-2p transition. The grey oblique line is the expected detuning considering $\omega_{23} = 155$ meV. d) Three selected spectra of the mid-IR pulse measured by a home-made Michelson interferometer. e) Rabi frequency as a function of $\sqrt{\phi_C}$ for a resonant mid-IR pump ($\omega_C = 155$ meV). The shadowed grey line follows the expected linear dependence.*

***Figure 3: Ultrafast screening of Rydberg-like excitonic levels.*** *a) Transient renormalization of the 1s (purple) and the 2p (pink) excitonic levels upon photoexcitation by $\omega_{pump}$. The inset shows three-pulse $\Delta T/T$ maps for selected delays $\Delta t$ measured with $\omega_C = 155$ meV. The respective ATS spectra are plotted with (black) and without (grey) the photoexcitation by $\omega_{pump}$. b) Sketch of the ultrafast screening of excitons with s- and p-symmetry wavefunctions, where $r$ indicates the exciton radius, i.e., the separation between a positively-charged hole (red circle at the origin of the axes) and a negatively-charged electron. In the plane, the effects of an above-gap pulsed excitation are sketched: i) a sudden photoinjection of free charges at $t = 0$; ii) the formation at later times of new bound excitons. The screening, which is mostly delocalized right after the photoexcitation, becomes more localized as exciton-exciton interactions are established.*

**ACKNOWLEDGMENTS**
The authors acknowledge insightful discussions with Atac Imamoğlu. This work was mainly supported by the European Research Council through the project INCEPT (grant agreement no. 677488). D.F., A.M., F.F., G.J. and E.M.R. acknowledge the support of the Gordon and Betty Moore Foundation through the grant CENTQC (no. GBMF12213). Monolayer sample preparation was supported by the Materials Science and Engineering Research Center (MRSEC) through NSF grant DMR-2011738 (XYZ). C.T. and G.C. acknowledge the European Union's Horizon Europe research and innovation program under the Marie Skłodowska-Curie PIONEER HORIZON-MSCA-2021-PF-GF (grant agreement no. 101066108). M.E. and F.V. were supported by the Cluster of Excellence "CUI: Advanced Imaging of Matter" of the Deutsche Forschungsgemeinschaft (DFG) – EXC 2056 – project ID 390715994.

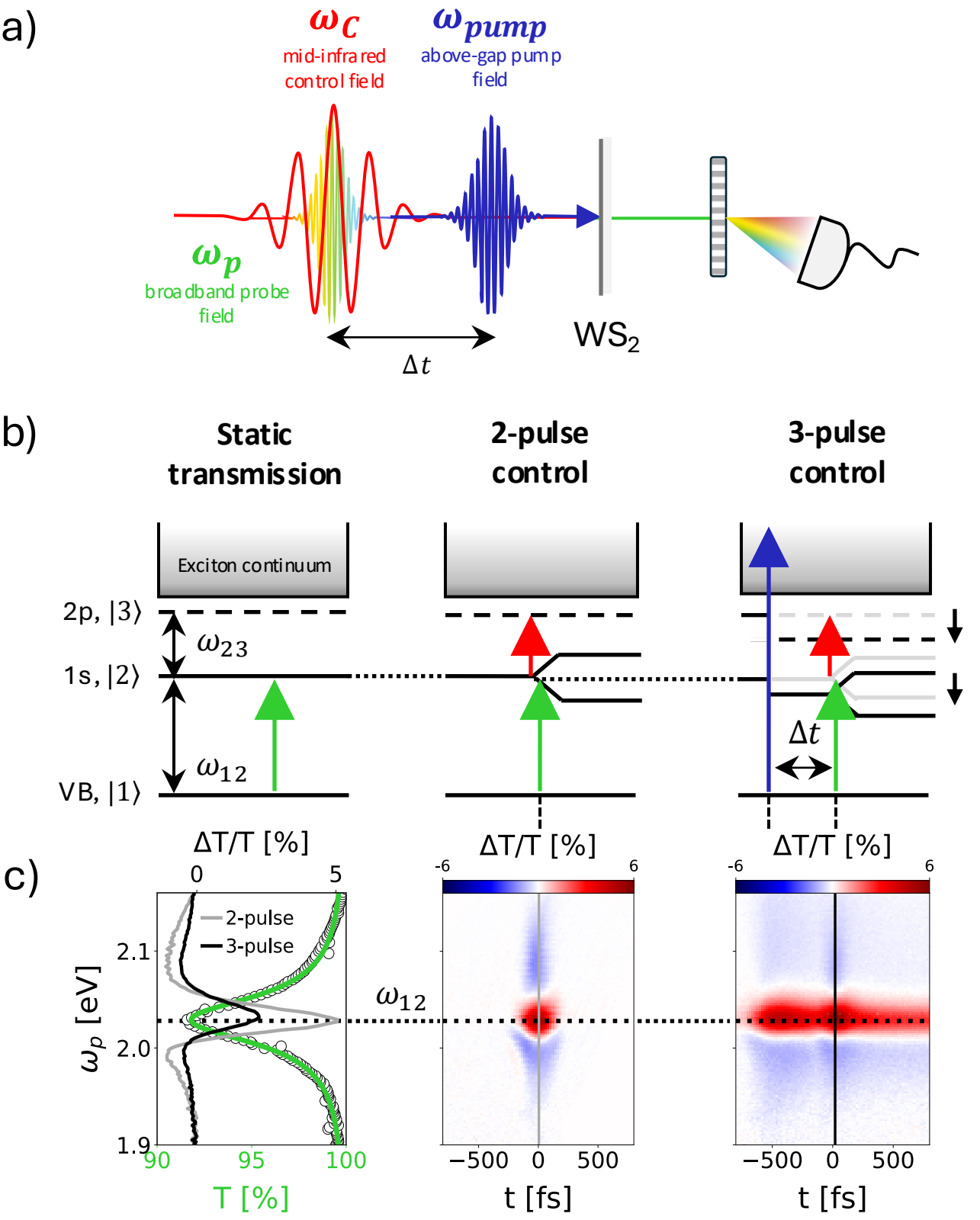
a)
$\omega_C$
mid-infrared control field
$\omega_{pump}$
above-gap pump field
$\omega_p$
broadband probe field
$\Delta t$
$WS_2$
b)
Static transmission
2-pulse control
3-pulse control
Exciton continuum
2p, |3⟩
1s, |2⟩
VB, |1⟩
$\omega_{23}$
$\omega_{12}$
$\Delta t$
c)
ΔT/T [%]
0
5
2-pulse
3-pulse
2.1
2.0
1.9
90
95
100
T [%]
$\omega_p$ [eV]
$\omega_{12}$
-6
6
-500
0
500
t [fs]

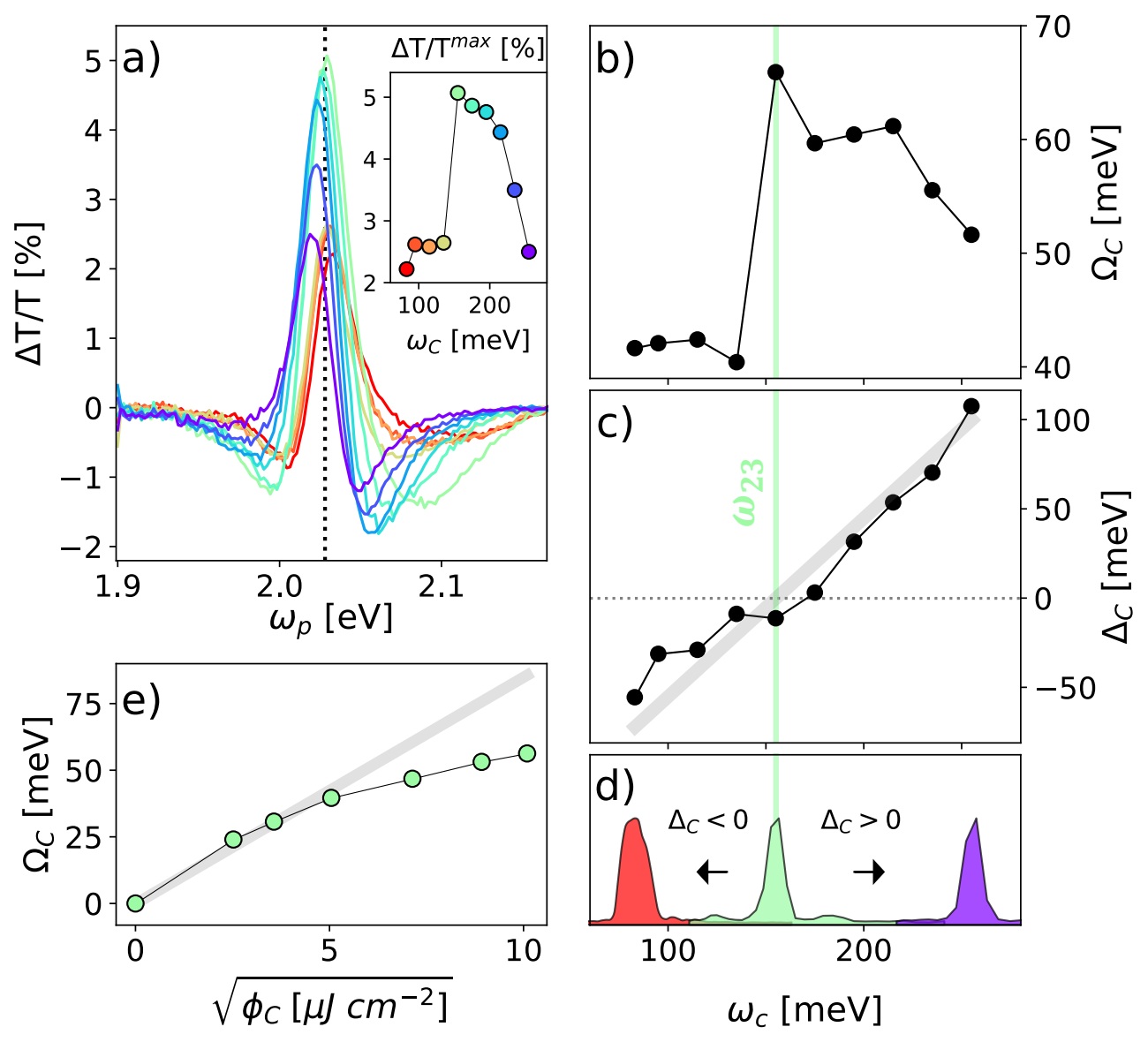

a)
ΔT/T [%]
$\omega_p$ [eV]
ΔT/T$^{max}$ [%]
$\omega_C$ [meV]
b)
$\Omega_C$ [meV]
c)
$\omega_{23}$
$\Delta_C$ [meV]
d)
$\Delta_C < 0$
$\Delta_C > 0$
$\omega_c$ [meV]
e)
$\Omega_C$ [meV]
$\sqrt{\phi_C\ [\mu J\ cm^{-2}]}$

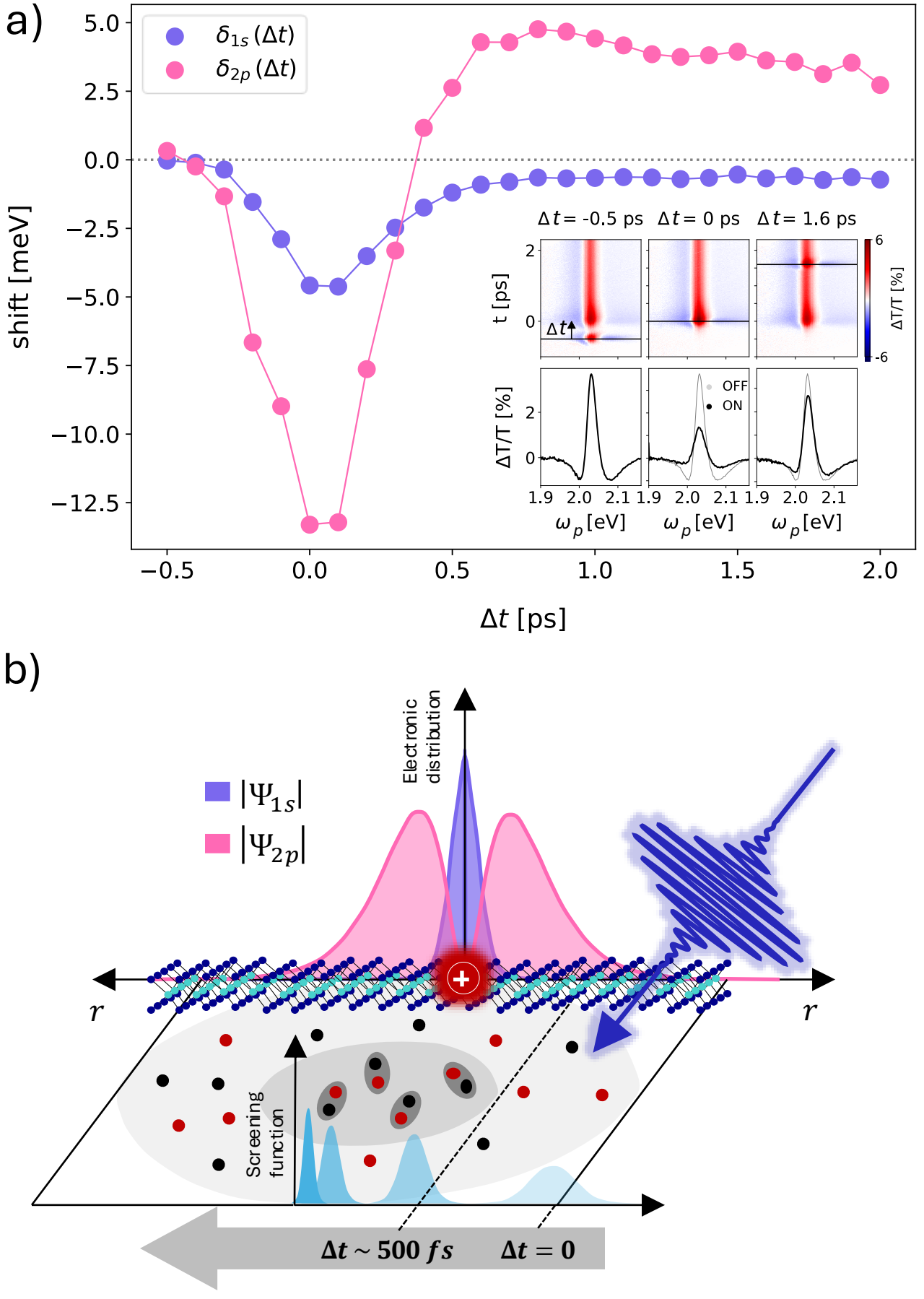

a)
$\delta_{1s}(\Delta t)$
$\delta_{2p}(\Delta t)$
shift [meV]
$\Delta t$ [ps]
$\Delta t$ = -0.5 ps
$\Delta t$ = 0 ps
$\Delta t$ = 1.6 ps
t [ps]
ΔT/T [%]
OFF
ON
$\omega_p$ [eV]
b)
Electronic distribution
$|\Psi_{1s}|$
$|\Psi_{2p}|$
r
Screening function
$\Delta t \sim 500\ fs$
$\Delta t = 0$

# Supplementary Materials for

# Tracking the Photoinduced Dynamics of a Dark Excitonic State in Single-Layer $WS_2$ via Resonant Autler-Townes Splitting

Angela Montanaro[1,2,3,*], Francesco Valiera[4,*], Francesca Giusti[1,2], Francesca Fassioli[3], Chiara Trovatello[5,6], Giacomo Jarc[1,2,3], Enrico Maria Rigoni[1,2,3], Fang Liu[7,8], Xiaoyang Zhu[7], Stefano Dal Conte[5], Giulio Cerullo[5,9], Martin Eckstein[4], and Daniele Fausti[1,2,3,†]

*1. Department of Physics, Università degli Studi di Trieste, Trieste, Italy.*
*2. Elettra Sincrotrone Trieste, Trieste, Italy.*
*3. Department of Physics, University of Erlangen-Nürnberg, Erlangen, Germany*
*4. Institute of Theoretical Physics, University of Hamburg, Hamburg, Germany*
*5. Dipartimento di Fisica, Politecnico di Milano, Milano, Italy*
*6. Department of Mechanical Engineering, Columbia University, New York, 10027, NY, USA*
*7. Department of Chemistry, Columbia University, New York, 10027, NY, USA*
*8. Department of Chemistry, Stanford University, Stanford, CA 94305, USA*
*9. Istituto di Fotonica e Nanotecnologie, Consiglio Nazionale delle Ricerche, Milano, Italy*

[*] These authors contributed equally to the work.
[†] Correspondence to: daniele.fausti@fau.de

**This PDF file includes:**

Supplementary Text (Notes from 1 to 9)
Supplementary Figures (Figs. from S1 to S15)

# Supplementary Note 1: Experimental setup

The source of the experimental apparatus was a regeneratively amplified Yb-KGW laser (Pharos, Light Conversion), which delivers 290 fs pulses at 1.2 eV with an average power of 20 W. The laser operated at a repetition rate of 10 kHz. The sample was excited by tunable mid-IR pulses (ranging from 83 to 255 meV) generated by difference frequency mixing in a GaSe crystal, from two near-infrared beams produced in a Twin Optical Parametric Amplifier (Orpheus TWIN by Light Conversion). A homemade Michelson interferometer, equipped with a mercury cadmium telluride detector, was used to measure the wavelength of the mid-IR pulses. A 3.1 eV pump pulse was obtained through second harmonic generation in a β-barium borate crystal, using the output from a Non-Collinear Parametric Amplifier (Orpheus-N by Light Conversion). Both pumps were mechanically modulated by two synchronized optical choppers running at 45 and 90 Hz. More details about the experimental setup are in Ref. [25].
The transmissivity of the sample under non-equilibrium conditions was probed using a broadband pulse covering 1.3 eV to 2.2 eV, obtained by white-light continuum generation of the Pharos laser output in a 6-mm-thick sapphire crystal. Prior to interaction with the sample, the probe was filtered using a bandpass filter (1.9–2.16 eV) to center its spectral window on the 1*s* bright exciton peak. When focused on the sample, the mid-IR pump, the blue pump and the probe beams have Gaussian spot sizes with FWHM of ~ 100 μm, ~ 100 μm and ~ 30 μm, respectively. All three pulses were linearly polarized.
The transmitted probe beam was diffracted by a transmission grating (1200 grooves/mm), and its spectral components were detected by a linear array of silicon photodiodes (NMOS by Hamamatsu), synchronized with the laser's repetition rate. To improve the signal-to-noise ratio, a reference beam (not interacting with the sample) was simultaneously measured and used to normalize the data, reducing fluctuations in the white-light probe. While no physical adjustments were made to correct for the temporal chirp of the broadband probe, all data were processed after acquisition to account for dispersion effects.
All the measurements discussed in the manuscript have been performed at room temperature.

# Supplementary Note 2: Sample preparation

Monolayers $WS_2$ are mechanically exfoliated from a bulk $WS_2$ single crystal (HQ Graphene) using a gold-tape exfoliation technique previously reported in Ref. [26]. The process begins by depositing a gold layer onto a clean Si wafer, followed by spin-coating with a polyvinylpyrrolidone (PVP) polymer layer. The PVP/Au stack is then lifted from the Si wafer using thermal release tape (Nitto) and brought into contact with a freshly cleaved $WS_2$ bulk crystal, exfoliating a large $WS_2$ monolayer. The tape/PVP/Au/$WS_2$ stack is then transferred onto a double-side-polished fused silica substrate. The thermal release tape is removed by heating to 100°C, and the PVP layer is eliminated through water rinsing and $O_2$ plasma treatment. Finally, the gold layer is dissolved using a KI/$I_2$ etchant solution (KI 99.9%, $I_2$ 99.99%, Alfa Aesar) in deionized water. After thorough rinsing with water and isopropanol, a clean $WS_2$ monolayer is obtained on the fused silica substrate.

# Supplementary Note 3: Theoretical model and simulations

The system has been described as an ensemble of independent 3-level systems, following the approach in Ref. [32]. Contrarily to Ref. [32], we used a ladder-type 3-level-system instead of a lambda-type – namely, the forbidden transition in our case is the one from the lowest energy level to the highest energy one, instead of that from lowest to intermediate.
The exciton dynamics has been described by the master equation in the interaction picture:

$$\dot{\rho} = -\frac{i}{\hbar}[H_0 + H_{Rabi}(t), \rho] + \mathcal{D}[\rho] \qquad \text{[Eq. S1]}$$

where $H_0$ is the unperturbed Hamiltonian of the 3-level-system

$$H_0 = \omega_1 \sigma_{11} + \omega_2 \sigma_{22}$$

with $\sigma_{ij} = |i\rangle\langle j|$, while $H_{Rabi}$ is the time-dependent Rabi Hamiltonian in Eq. 1 in the main text, where the Rabi frequencies are defined as

$$\Omega_p = |\vec{\mu}_{12}^* \cdot \vec{E}_p|, \qquad \Omega_C = |\vec{\mu}_{23}^* \cdot \vec{E}_C|,$$

with $\mu_{ij}$ being the dipole matrix element of the transition between the states $|i\rangle$ and $|j\rangle$, and $\vec{E}_p, \vec{E}_C$ the electric field amplitudes of the probe and control pulses. Moreover, the dissipator operator $\mathcal{D}$ describes decay processes from higher- to lower-energy levels and the dephasing of the off-diagonal density matrix elements; it is defined as

$$\mathcal{D}[\rho] = \frac{\Gamma_{31}}{2}(2\sigma_{12}\rho\sigma_{21} - \{\sigma_{22}, \rho\}) + \frac{\Gamma_{32}}{2}(2\sigma_{23}\rho\sigma_{32} - \{\sigma_{33}, \rho\}) + \frac{\gamma_{2d}}{2}(2\sigma_{22}\rho\sigma_{22} - \{\sigma_{22}, \rho\}) + \frac{\gamma_{3d}}{2}(2\sigma_{33}\rho\sigma_{33} - \{\sigma_{33}, \rho\}).$$

The coefficients $\Gamma_{31}$, $\Gamma_{32}$, $\gamma_{2d}$ and $\gamma_{3d}$ are phenomenological: the former two respectively quantify the decay rate from $|2\rangle$ to $|1\rangle$ and from $|3\rangle$ to $|2\rangle$; whereas the latter two respectively describe the dephasing processes of the level $|2\rangle$ and $|3\rangle$, which effectively reduce the coherence of the system. Notice that the presence of the dephasing terms is critical for tuning the linear optical susceptibility to the right shape and eventually provides the knob to tune the amplitude of the transparency window.

Let now $\rho_{ij} = \langle i|\rho|j\rangle$, $i, j = 1,2,3$ be the generic matrix element of the density matrix. The values of the diagonal terms $\rho_{11,}\ \rho_{22,}\ \rho_{33}$ reflect the populations of the states $|1\rangle, |2\rangle, |3\rangle$. For the off-diagonal terms one can switch to a rotating frame by setting

$$\rho_{12} = r_{12} e^{i\omega_p t},$$
$$\rho_{23} = r_{23} e^{i\omega_C t},$$
$$\rho_{13} = r_{13} e^{i(\omega_p + \omega_C)t}.$$

The Bloch equations then follow from the master equation in Eq. S1:

$$\dot{\rho}_{11} = -\frac{i}{2}(\Omega_p r_{12} - \Omega_p^* r_{12}^*) + \Gamma_{12}\rho_{22};$$
$$\dot{\rho}_{22} = \frac{i}{2}(\Omega_p r_{12} - \Omega_p^* r_{12}^*) - \frac{i}{2}(\Omega_c r_{23} - \Omega_c^* r_{23}^*) - \Gamma_{21}\rho_{22} + \Gamma_{32}\rho_{33};$$

$$\dot{\rho}_{33} = \frac{i}{2}(\Omega_c r_{23} - \Omega_c^* r_{23}^*) - \Gamma_{32}\rho_{33};$$
$$\dot{r}_{12} = \frac{i}{2}\Omega_p^*(\rho_{22} - \rho_{11}) - \frac{i}{2}\Omega_c r_{13} - (i\Delta_p + \gamma_{12})r_{12};$$
$$\dot{r}_{13} = -\frac{i}{2}\Omega_c^* r_{12} + \frac{i}{2}\Omega_p^* r_{23} - (i(\Delta_p + \Delta_C) + \gamma_{13})r_{13};$$
$$\dot{r}_{23} = \frac{i}{2}\Omega_p r_{13} + \frac{i}{2}\Omega_c^*(\rho_{33} - \rho_{22}) - (i\Delta_C + \gamma_{23})r_{23}.$$

The new decay rates have been introduced

$$\gamma_{12} = \frac{1}{2}(\Gamma_{21} + \gamma_{2d}), \qquad \gamma_{13} = \frac{1}{2}(\Gamma_{32} + \gamma_{3d}), \qquad \gamma_{23} = \gamma_{12} + \gamma_{23},$$

respectively quantifying the lifetime of the off-diagonal matrix elements $r_{ij}$.

The term $r_{12}$ in particular is needed for computing the contribution to the optical susceptibility of the transition $|1\rangle$-$|2\rangle$, which is the one measured in the experiment:

$$\chi(\Delta_p, \Delta_C) = C\frac{r_{12}^*(\Delta_p, \Delta_C)}{\Omega_p}$$

[Eq. S2]

where

$$C = \frac{n\mu_{12}^2}{2\epsilon_0}.$$

The quantity $n$ is an effective density of excitons per unit volume, while $\mu_{12}$ is the dipole moment of the transition. The former can be thought of as the density of available excitonic states of the systems, while the actual population of the states $|i\rangle$ will be given by $n\rho_{ii}$.

The equilibrium value of $\chi$ is determined by solving the equations of motion for $r_{12}$ in the stationary case, namely when $\dot{r}_{ij} = 0\ \forall i, j$. It is reasonable to assume that the intensity of the probe field is much smaller than that of the control field, therefore that the Rabi frequencies would satisfy

$$\lambda = |\Omega_p|/|\Omega_c| \ll 1;$$

Furthermore, this implies that the states $|2\rangle$ and $|3\rangle$ have negligible population, thus one can also assume that $\rho_{11} \approx 1$, while $\rho_{22} \approx \rho_{33} \approx 0$.

With such approximations, in the stationary regime one finds the following linear system for the coefficients $r_{ij}$:

$$\begin{pmatrix} -(i\Delta_p + \gamma_{12}) & -i\Omega_C / 2 & 0 \\ -i\Omega_C / 2 & -[i(\Delta_p + \Delta_C) + \gamma_{13}] & i\Omega_p / 2 \\ 0 & i\Omega_p / 2 & -(i\Delta_C + \gamma_{23}) \end{pmatrix} \begin{pmatrix} r_{12} \\ r_{13} \\ r_{23} \end{pmatrix} = \begin{pmatrix} i\Omega_p^* / 2 \\ 0 \\ 0 \end{pmatrix},$$

which can then be solved in perturbation theory to first order in $\lambda$. Inserting the value of $r_{12}$ into the expression for $\chi$ in Eq. S2 finally leads to an expression for the susceptibility

$$\chi(\Delta_p) = \frac{C}{\left|\frac{\Omega_C^2}{4} + (i\Delta_p + \gamma_{12})(i\Delta + \gamma_{13})\right|^2}\left[\left(\frac{\Omega_C^2}{4}\Delta - \Delta_1(\gamma_{13}^2 + \Delta^2)\right) + i\left(\frac{\Omega_C^2}{4}\gamma_{13} + \gamma_{12}(\gamma_{13}^2 + \Delta^2)\right)\right]$$

[Eq. S3]

where we have defined $\Delta = \Delta_p + \Delta_C$.

Finally, the transmission of the monolayer has been worked out within the thin-film approximation. The approximation considers the current density in 1L-$WS_2$ to have a delta-like profile and takes into account the influence of the semi-infinite (500μm thick) dielectric substrate of fused silica [20]. By solving the Maxwell's equations with the appropriate boundary conditions, it is possible to obtain an expression for the transmittance of the monolayer:

$$T(\omega_p) = \frac{4n_{sub}}{\left(1 + n_{sub} + \frac{\omega_p d}{c}\Im(\chi)\right)^2 + \left(\frac{\omega_p d}{c}\Re(\chi)\right)^2}$$

[Eq. S4]

where $n_{sub}$ is the refractive index of the substrate ($n_{sub} = 1.458$ for fused silica at 2 eV) and $d$ is the thickness of the monolayer ($d = 0.67\ nm$ for 1L-$WS_2$). Eq. S4 has been used to fit the experimental data.

We outline now some features of the model and discuss numerical simulations performed across the parameter space.
Firstly, we point out that the expression obtained for $\chi$ (Eq. S3) predicts electromagnetically induced transparency (EIT), namely the opening of a narrow transparency window in the absorption spectrum, when the following conditions are satisfied:

$$\gamma_{12} \gg |\Omega_c| \gg \sqrt{\gamma_{12}\gamma_{13}}.$$

The rate $\gamma_{12}$ provides then a natural energy scale to which compare other energy quantities.

Let us first consider the case $\gamma_{13} = 0$, namely when there is not any additional dephasing. In this situation, the transparency is perfect, meaning that the imaginary part of the susceptibility is exactly zero at zero probe detuning ($\Delta_p = 0$). This is clear in the light blue curve in Fig. S1a. However, when the Rabi frequency of the control field $\Omega_c$ is increased towards values comparable to or higher than $\gamma_{12}$, the shape of the susceptibility changes qualitatively: instead of a sharp and narrow transparency window, a line splitting occurs.

A finite dephasing $\gamma_{13} \neq 0$ quenches the transparency, as it is illustrated in Fig. S1b. As $\gamma_{13}$ increases at a fixed control field amplitude, the imaginary part of the susceptibility at zero probe detuning takes a finite value and it is completely canceled out when $\gamma_{13} > \gamma_{12}$. In the Autler-Townes regime (Fig. S1c) – i.e., for more intense control fields – the splitting is also smeared out by increasing $\gamma_{13}$.

Finally, in Fig. S1d the effect of the detuning of the control field is highlighted. The choice of parameters $\gamma_{13}$ and $\Omega_c$ is such that the system is in the EIT regime, but we highlight that a similar trend is present also in the Autler-Townes regime. The detuning shifts the position of the transparency window, making the susceptibility profile asymmetric.

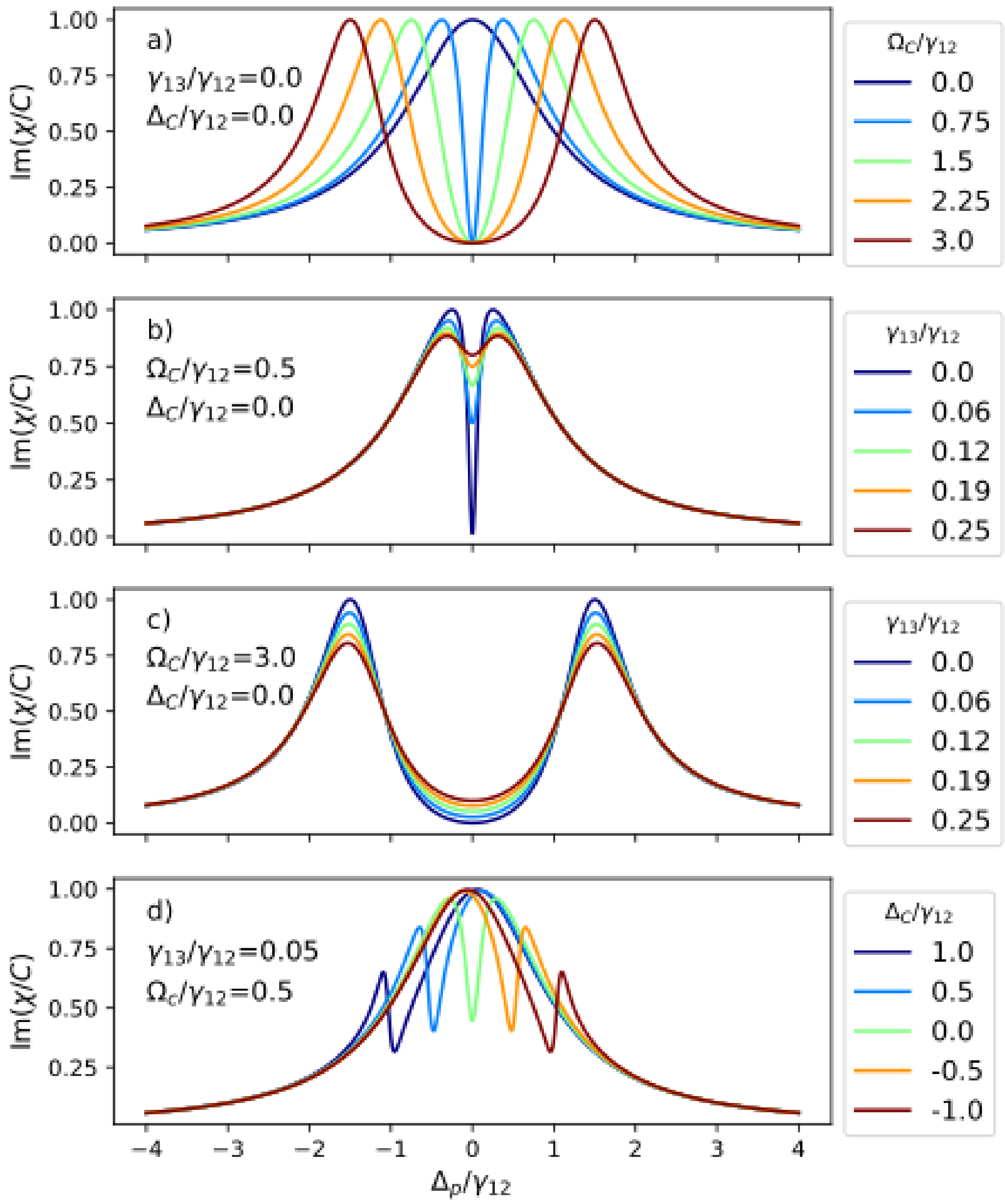

*Figure S1:* ***Imaginary part of linear susceptibility as a function of different parameters.*** *a) Dependence of the susceptibility on the Rabi frequency of the control field. At low values of $\Omega_C$ ($0.25\gamma_{12}$, $0.5\gamma_{12}$) the transparency window is very narrow, indicating that EIT occurs; as the Rabi frequency is increased, a splitting of the absorption spectrum takes place. b) Effect of dephasing coefficient $\gamma_{13}$ in the EIT regime and in c) in the Autler-Townes one. d) Effect of finite pump detuning (notice that since $\gamma_{13}$ is finite the EIT is not perfect).*

# Supplementary Note 4: Analysis of the static absorption

The static absorption of the sample has been measured at room temperature using the same broadband probe employed in the time-resolved experiments.
Three different spectra have been acquired:

i. the dark spectrum of the photodiode array ($S_{bkg}$).
ii. the spectrum transmitted by the fused silica substrate ($S_{sub}$).
iii. the spectrum transmitted by the sample ($S_{sam}$).

In Fig. S2a we plot the spectra transmitted by the substrate (grey) and the sample (black) after subtracting the background. We observe an overall decrease of the amount of light transmitted by the sample, and in particular a dip at around 2 eV. This reduced transmission corresponds to the absorption by the bright 1$s$ exciton.

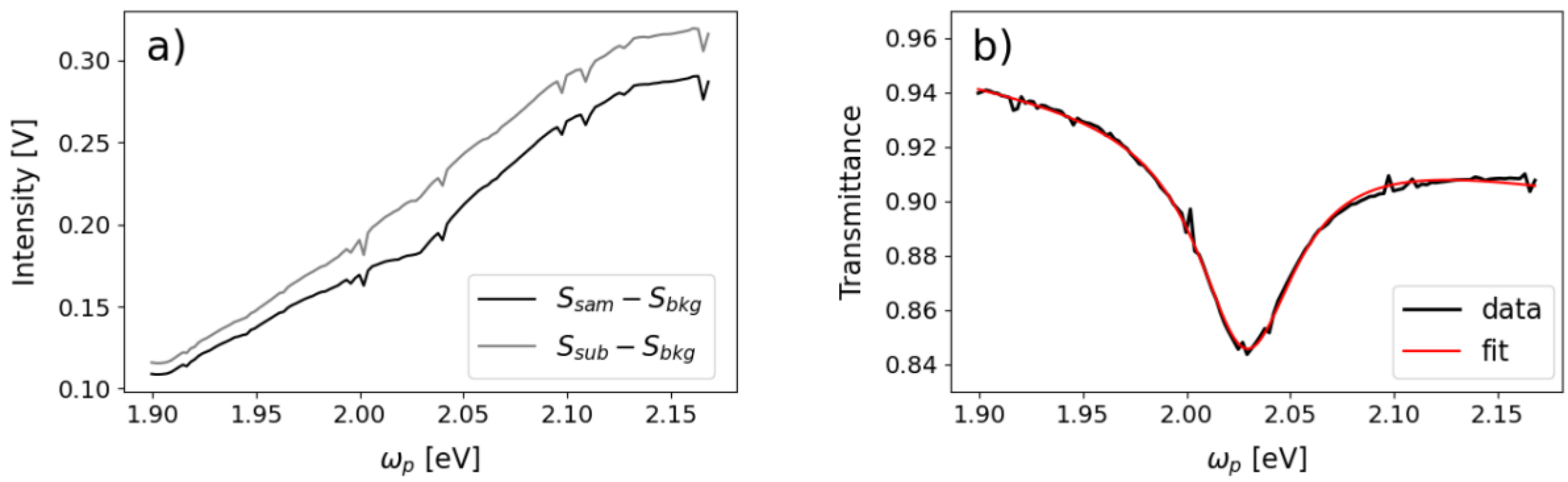

*Figure S2:* ***Static absorption of 1L-WS₂.*** *a) Spectra transmitted by the substrate (grey) and the sample (black) at equilibrium. b) Transmittance of the sample fitted using Eq. S4 and considering the expression for the equilibrium susceptibility (Eq. S5).*

We plot in Fig. S2b the transmittance of the sample ($T$), obtained as ratio between the curves in Fig. S2a. The 1$s$ excitonic peak features an absorption of ~8% and has a Lorentzian shape centered at 2.029 eV - in agreement with previous studies [27] - superimposed to a linear background. In the spectrum plotted in Fig. 1c (green line), the linear background has been subtracted.

To fit the data, we used the model developed for the complex susceptibility (Eq. S3) in the limit in which the system is at equilibrium – i.e., no mid-infrared control field is applied. In this condition, Eq. S3 can be rearranged considering that $\Omega_C = 0$. By separating the real and imaginary part, we obtain:

$$\begin{cases} \Re[\chi_{eq}(\omega_p)] = -C\dfrac{\omega_p - \omega_{12}}{(\omega_p - \omega_{12})^2 + \gamma_{12}^2} \\ \Im[\chi_{eq}(\omega_p)] = +C\dfrac{\gamma_{12}}{(\omega_p - \omega_{12})^2 + \gamma_{12}^2} \end{cases}$$

[Eq. S5]

The transmittance at equilibrium has been then fitted by Eq. S4 using the equilibrium susceptibility in Eq. S5. The fitting function has therefore five free parameters: $C$, $\omega_{12}$, $\gamma_{12}$ and two remaining parameters for the linear background.
The static absorption from 1$s$ exciton is found centered at $\omega_{12} = 2.029$ eV and its width is $\gamma_{12} = 21$ meV. Importantly, consistent values are obtained by fitting the absorption curve with a Lorentzian function, providing a valuable sanity-check for our model. Furthermore, the fitted value of the constant $C = 0.455\ eV$ provides an estimate for the dipole moment $\mu_{12} = 45$ D, which agrees well with previous works [33].

# Supplementary Note 5: Fits of the induced Autler-Townes splitting

## i. As a function of mid-infrared detuning

To study the dependence of the Autler-Townes splitting on the detuning of the control field, we changed the mid-IR photon energy in the range 83-255 meV and we measured the transient transmissivity. The fluence was kept constant and equal to 50 $\mu J cm^{-2}$ throughout the measurements, while the fluence of the probe pulse was maintained to 2 $\mu J cm^{-2}$. This justifies the approximation on the Rabi frequencies made in the model, i.e., $\Omega_p \ll \Omega_C$.
The maps in Fig. S3 show time- and energy-resolved $\Delta T/T$ maps measured. The color scale represents the pump-induced change in the probe transmissivity $\Delta T/T$, which in the perturbative limit corresponds to the opposite of the transient absorption $\Delta A$. This means that positive values in the maps (red) correspond to a decrease in the probe absorption. The map acquired at $\omega_C = 155$ meV is the one reported in Fig. 1c (middle panel) and corresponds to the resonance with the $1s$-$2p$ transition ($\Delta_C = 0$). In this case, as discussed in the main text, the transparency window at the temporal overlap ($t = 0$) features a symmetric shape across the energy axis with a negative-positive-negative profile. When the mid-IR photon energy is decreased to the minimum value ($\omega_C = 83$ meV, $\Delta_C < 0$), the transparency has a reduced intensity and an asymmetric shape, with a more pronounced negative shoulder at lower probe energies. The asymmetry is reversed for the highest mid-IR photon energy measured ($\omega_C = 255$ meV, $\Delta_C > 0$).

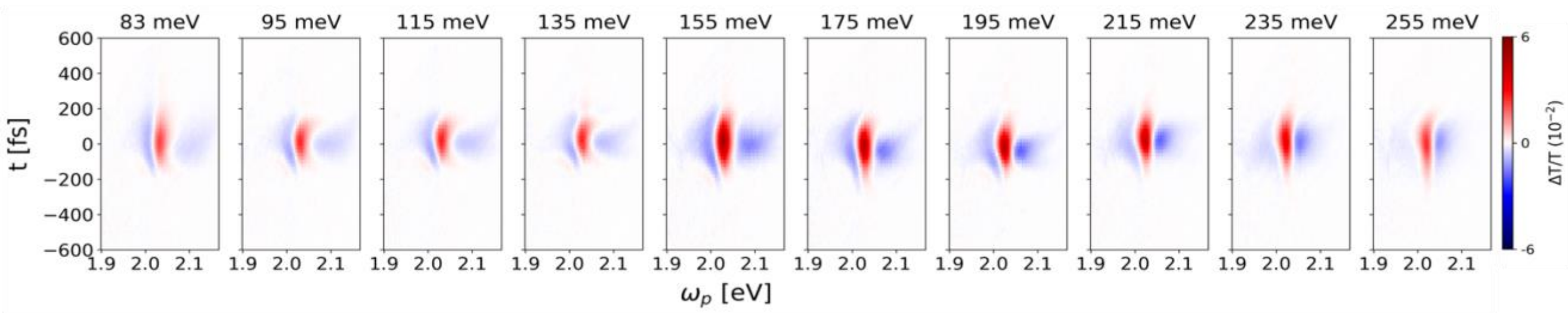

*Figure S3: **ATS splitting as a function of mid-IR deteuning.** Time- and energy-resolved $\Delta T/T$ maps upon interaction with mid-infrared fields of varying photon energy.*

This trend is even clearer in Fig. S4, where the black dots show horizontal cuts of the maps in Fig. S3 at the temporal overlap. All the curves have been globally fitted using the following expression:

$$\frac{\Delta T}{T} = \frac{T(\omega_p) - T(\omega_p, \Omega_C = 0)}{T(\omega_p, \Omega_C = 0)}$$

[Eq. S6]

where $T(\omega_p)$ is Eq. S4 accounting for the transmission when the mid-IR control field impinges on the sample and $T(\omega_p, \Omega_C = 0)$ is again Eq. S4 but with a null Rabi frequency, that is including the equilibrium susceptibility (Eq. S5).
The global fitting procedure has been carried out using the *lmfit* routine in Python. The expression of $T(\omega_p)$ has five parameters: $C$, $\omega_{12}$, $\Omega_C$, $\Delta_C$, $\gamma_{12}$ and $\gamma_{13}$. Among these, $C$, $\omega_{12}$ and $\gamma_{12}$ are fixed parameters and their values are constrained by the equilibrium measurements (Supplementary Note 4). Furthermore, being $\gamma_{13}$ a material-dependent property, it should be the same across all the measurements performed by detuning the mid-IR control field. We therefore constrained the global fit in such a way that $\gamma_{13}$ is a free fitting parameter but it should have the same value for all the curves in

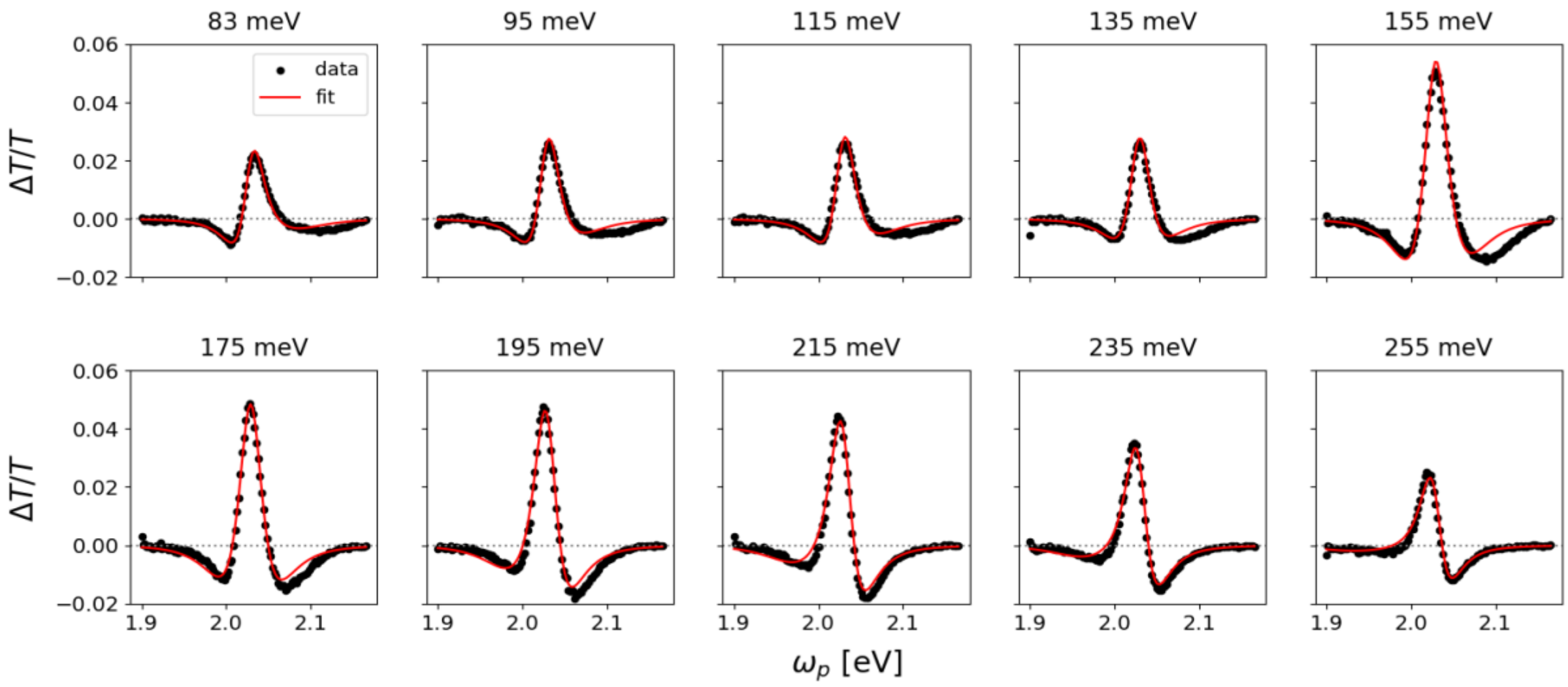

*Figure S4: **Comparison of experimental and fitted spectra as a function of mid-IR detuning.** Spectra (black dots) at the temporal overlap taken from the maps in Fig. S3 and corresponding fits (red lines).*

Fig. S4. The parameters $\Omega_C$ and $\Delta_C$ are instead free to change. The results of the global fit are plotted as red lines superimposed to the data in Fig. S4.
As discussed in the main text, the best global fit is obtained for $\gamma_{13} = 65$ meV. $\Omega_C$ and $\Delta_C$ extracted from the fit are instead plotted as a function of the mid-IR photon energy in Fig. 2b and 2c, respectively. In particular, the dependence of $\Delta_C$ on $\omega_C$ constitutes a "sanity-check" for the model: as expected, the $\Delta_C = 0$ condition is obtained for $\omega_C = 155$ meV. Both the negative and positive detunings are correctly retrieved.
In order to study how the 1$s$ absorption is modified by the interaction with the mid-IR control field, we plot in Fig. S5 the imaginary part of the susceptibility in Eq. S3. For each colored curve in Fig. S5, we included in the calculation the fitting parameters obtained by the global fits of the curves in Fig. S4. The black line is instead the equilibrium absorption in the absence of the mid-IR field ($\Omega_C = 0$). The plot indicates that the strongest quench of the absorption is observed at $\omega_C = 155$ meV (light green curve). However, a more pronounced splitting of the Autler-Townes doublet (like the one simulated in Fig. S1a) is never achieved within the parameter space explored in the experiment.

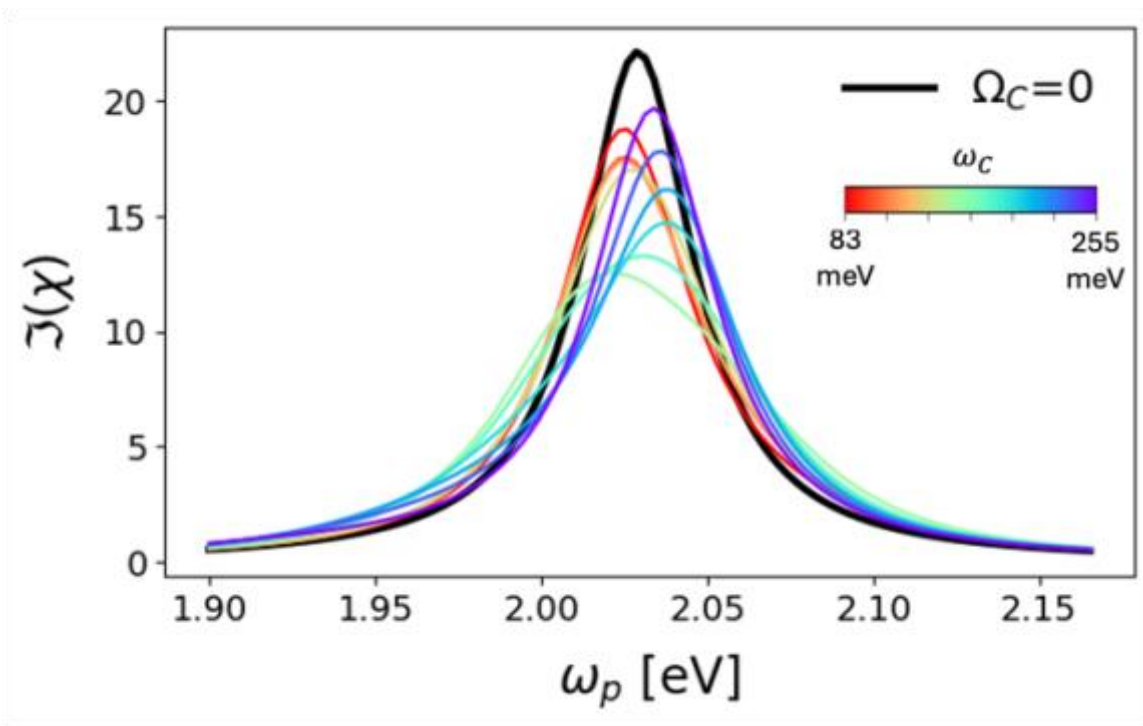

*Figure S5: **Calculated $\Im(\chi)$ as a function of mid-IR deteuning.** Imaginary part of the susceptibility calculated in the absence of the mid-infrared field ($\Omega_C = 0$) and upon intetaction with a mid-infrared field having the parameters obtained by fitting the experimental curves in Fig. S4.*

## ii. As a function of mid-infrared fluence

The dependence of the Autler-Townes splitting on the control field intensity has been studied by keeping fixed the photon energy of the mid-IR field on resonance with the 1$s$-2$p$ transition ($\omega_C = 155$ meV) and changing its fluence.

Fig. S6 shows the transient $\Delta T/T$ maps for the set of fluences measured. The spectra at the temporal overlap are plotted in Fig. S7 (black dotted curves). We fitted the spectra following the same approach discussed in the previous paragraph. In this case, the only parameter let free to change for every curve was $\Omega_C$. The parameters $\Delta_C$ and $\gamma_{13}$ have been instead constrained so that they have the same value across the dataset. The optimal fit has been achieved for $\Delta_C = 14$ meV and $\gamma_{13} = 65$ meV. The slightly positive detuning obtained is visible in the red fitting curves in Fig. S7, which feature an asymmetric shape especially at higher fluences. The decay rate $\gamma_{13}$ is then consistent with the one found in the previous set. The Rabi frequency extracted is plotted in Fig. 3b of the main text as a function of the mid-IR fluence. The imaginary part of the susceptibility calculated for each fluence is plotted in Fig. S8. A complete splitting of the doublet is not observed even at the highest fluence measured.

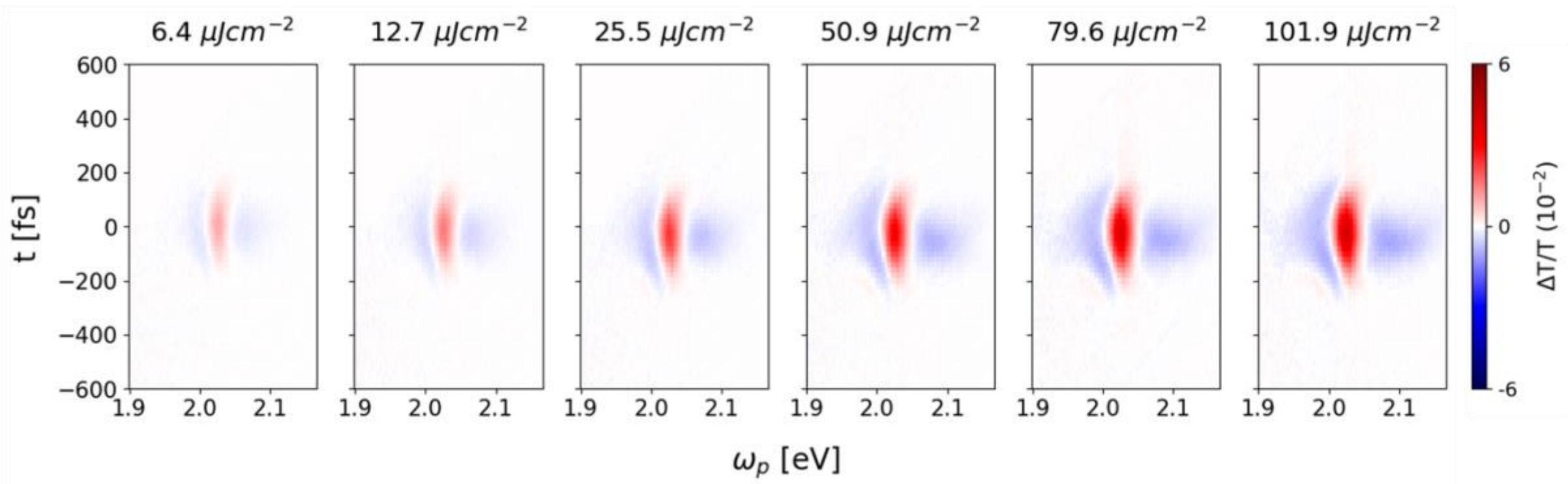

*Figure S6:* ***ATS splitting as a function of mid-IR fluence.*** *Time- and energy-resolved ΔT/T maps upon interaction with mid-infrared fields of varying fluence at constant photon energy ($\omega_C = 155\ meV$).*

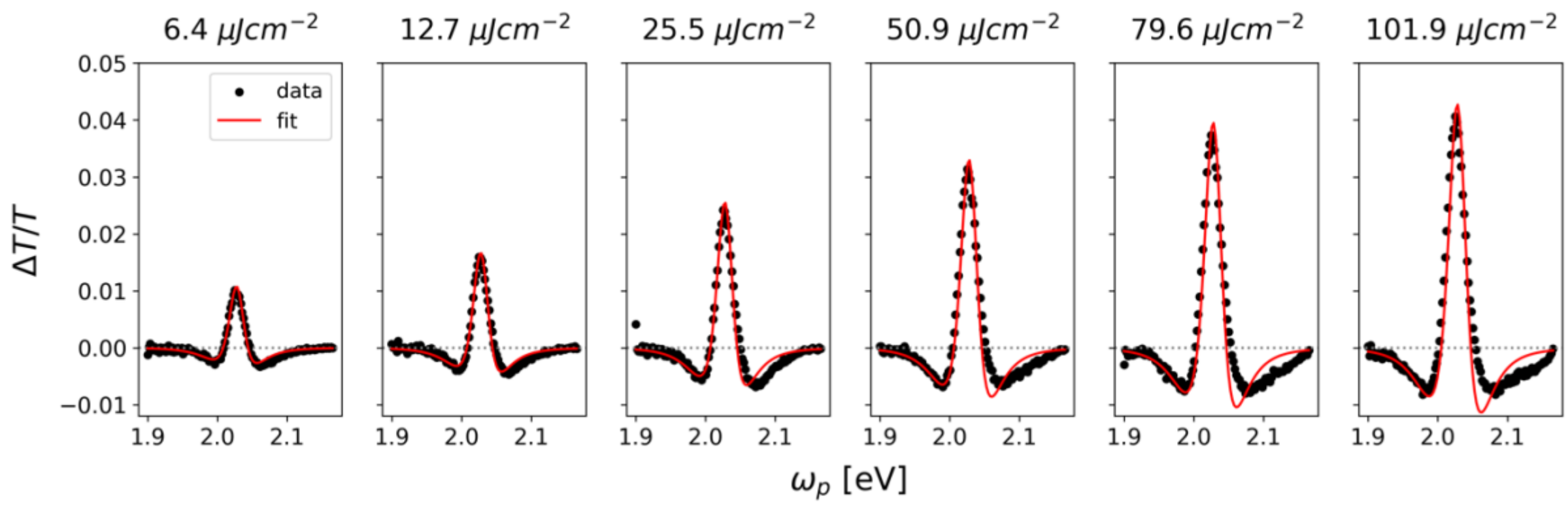

*Figure S7:* ***Comparison of experimental and fitted spectra as a function of mid-IR fluence.*** *Spectra (black dots) at the temporal overlap taken from the maps in Fig. S6 and corresponding fits (red lines).*

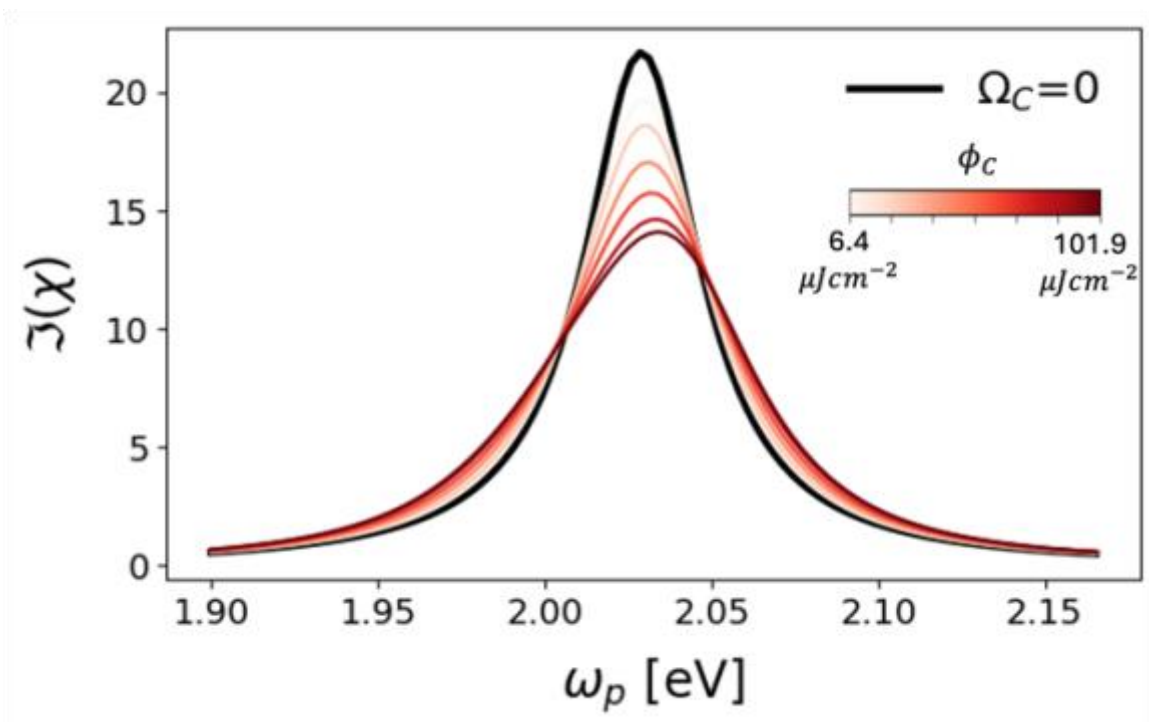

*Figure S8:* ***Calculated $\Im(\chi)$ as a function of mid-IR fluence.*** *Imaginary part of the susceptibility calculated in the absence of the mid-infrared field ($\Omega_C = 0$, black line) and upon intetaction with a mid-infrared field with the parameters obtained by fitting the experimental curves in Fig. S7.*

# Supplementary Note 6: Fluence dependence for detuned mid-IR driving

We measured the transient $\Delta T/T$ maps as a function of fluence (as in Fig. S6) for other two selected photon energies of the mid-IR control field, $\omega_C = 83$ meV and $\omega_C = 275$ meV. These two energies correspond to a negative and a positive detuning, respectively. The fluence-dependent curves have been fitted following the same procedure described in the previous section. The two independent sets have been globally fitted with the constraint that $\gamma_{13}$ is the same for all curves and that $\Delta_C$ is the same for all the fluence-dependent curves within the same set. The extrapolated value of $\gamma_{13}$ was again 65 meV, confirming the robustness of the fitting procedure. We retrieved a value of $\Delta_C = -44$ meV and $\Delta_C = 124$ meV for the negative and positive detuning, respectively.

In Fig. S9 we plot the Rabi frequencies extrapolated from the fits as a function of the square root of the mid-IR fluence for both sets. The shadowed grey lines follow a linear dependence and are a guide for the eye. Both curves are compatible with a linear trend and do not display saturation at higher fluences. This is consistent with the expected trend of the Rabi frequency of the mid-IR pulse which should linearly scale with the amplitude of the field. The saturated behavior observed at $\omega_C = 155$ meV (Fig. 2e) is thus peculiar of an excitation by a resonant field.

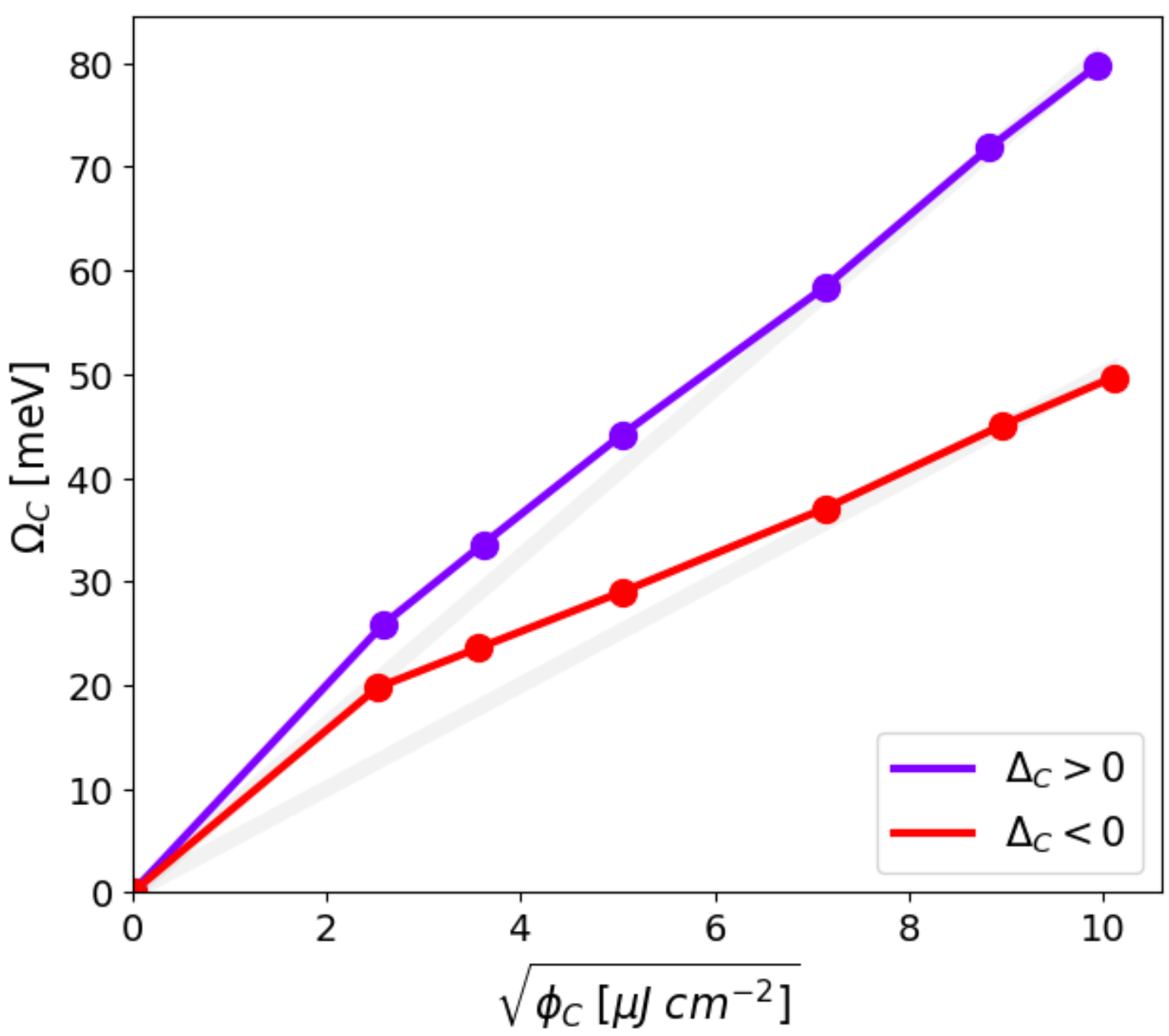

*Figure S9:* ***Rabi splitting as a funtion of mid-IR fluence.*** *Rabi frequency as a function of the square root of the mid-IR fluence for a positively- (purple, $\omega_C = 275$ meV) and a negatively-detuned (red, $\omega_C = 83$ meV) mid-IR pump as extrapolated from the global fits.*

# Supplementary Note 7: Fit of the above-gap pump-induced dynamics

Fig. S10a shows the $\Delta T/T$ map measured upon photoexcitation by an above-gap pump at $\omega_{pump} = 3.1$ eV at fluence 40 $\mu J cm^{-2}$. At the temporal overlap, the signal presents a second-derivative-like profile centered at the energy of the $1s$ exciton at equilibrium, denoted by the dotted horizontal line. At later times ($t > 500$ fs), the signal exhibits a derivative-like shape, displaying a negative shoulder at lower energies. $\Delta T/T$ spectra at $t = 0$ and $t = 1$ ps are plotted in the inset as solid and dashed lines, respectively.
To extract the temporal evolution of the central frequency ($\omega_{12}$) and the width ($\gamma_{12}$) of the 1s transition, we performed a differential fit - based on the absorption spectrum measured at equilibrium - at every time delay measured. The black dotted curves in Fig. S11 are representative $\Delta T/T$ spectra at selected times. The fit function used is:

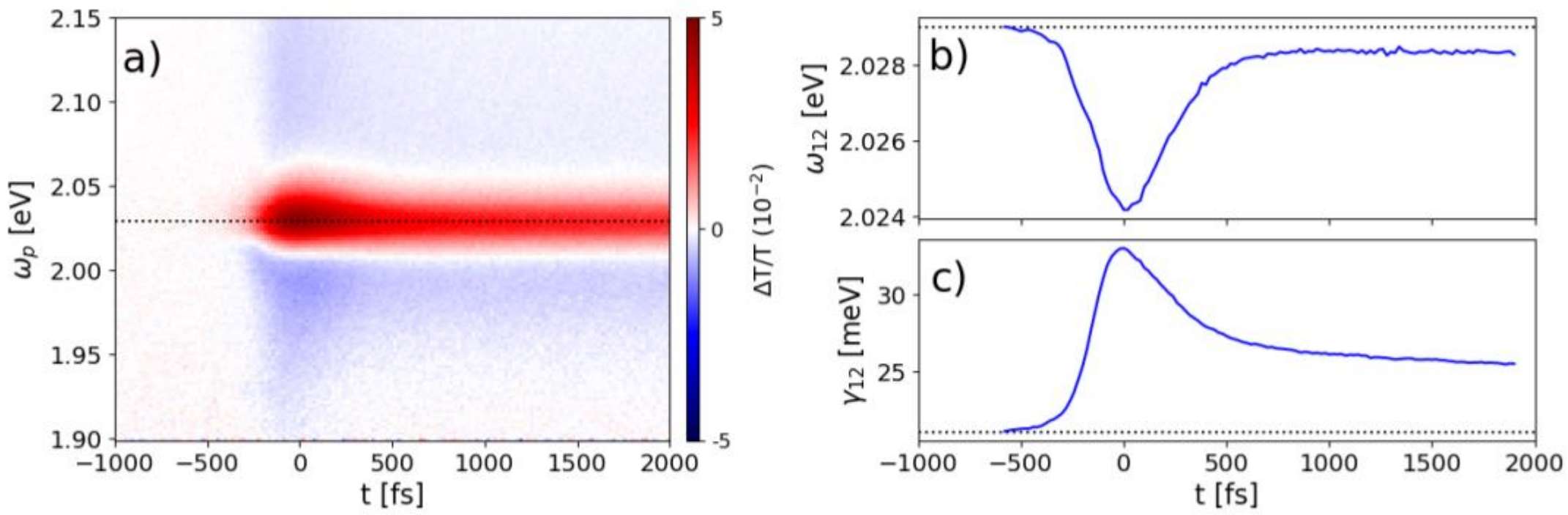

*Figure S10:* ***Pump-induced dynamics of the bright 1s exciton.*** *a) Transient transmissivity map as function of probe photon energy and time delay, after a photoexcitation at $\omega_{pump} = 3.1\ eV$. The dotted horizontal line indicates the energy of the bright 1s exciton measured at equilibrium. b) Temporal evolution of the 1s exciton energy after photoexcitation, as extrapolated from the differential fit. c) Temporal evolution of the linewidth of the 1s exciton.*

$$\frac{\Delta T}{T} = \frac{T^{exc}(\omega_p, \Omega_C = 0) - T(\omega_p, \Omega_C = 0)}{T(\omega_p, \Omega_C = 0)}$$

Where $T(\omega_p, \Omega_C = 0)$ is the transmittance in Eq. S4 evaluated at equilibrium (static absorption fit in Fig. S2b), while $T^{exc}(\omega_p, \Omega_C = 0)$ has the same functional form but the parameters $C$, $\omega_{12}$ and $\gamma_{12}$ are let free to vary. We find that the parameter $C$ does not significantly influence the fit, as the optimal values for the central frequency and the width of the transition remain unchanged whether $C$ is held constant or allowed to vary freely. The corresponding fits are overlaid as red lines in Fig. S11, while the temporal evolution of the parameters $\omega_{12}$ and $\gamma_{12}$ is plotted in Fig. S10a and S10b, respectively.

In Fig. S12, we plot the calculated transmittance based on the out-of-equilibrium parameters derived from the fitting procedure for several representative time delays. The black curve corresponds to the equilibrium transmittance prior to photoexcitation. The results demonstrate that the most significant perturbation occurs at temporal overlap ($t = 0$, dark blue line), and it is characterized by a redshift and a broadening of the absorption peak. As the time delay increases (light blue curve), the out-of-equilibrium transmittance gradually converges toward the static condition.

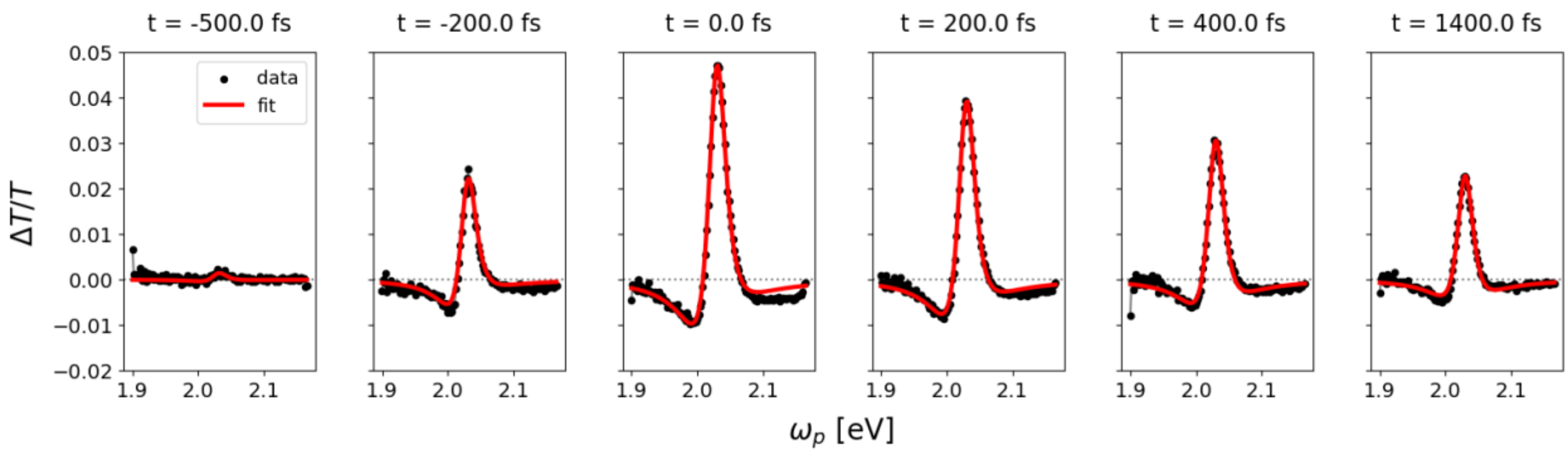

*Figure S11:* ***1s exciton spectra as a function of pump-probe delay.*** *ΔT/T spectra at selected time delays between the visible pump and the broadband probe (black dots) and corresponding fits (red lines).*

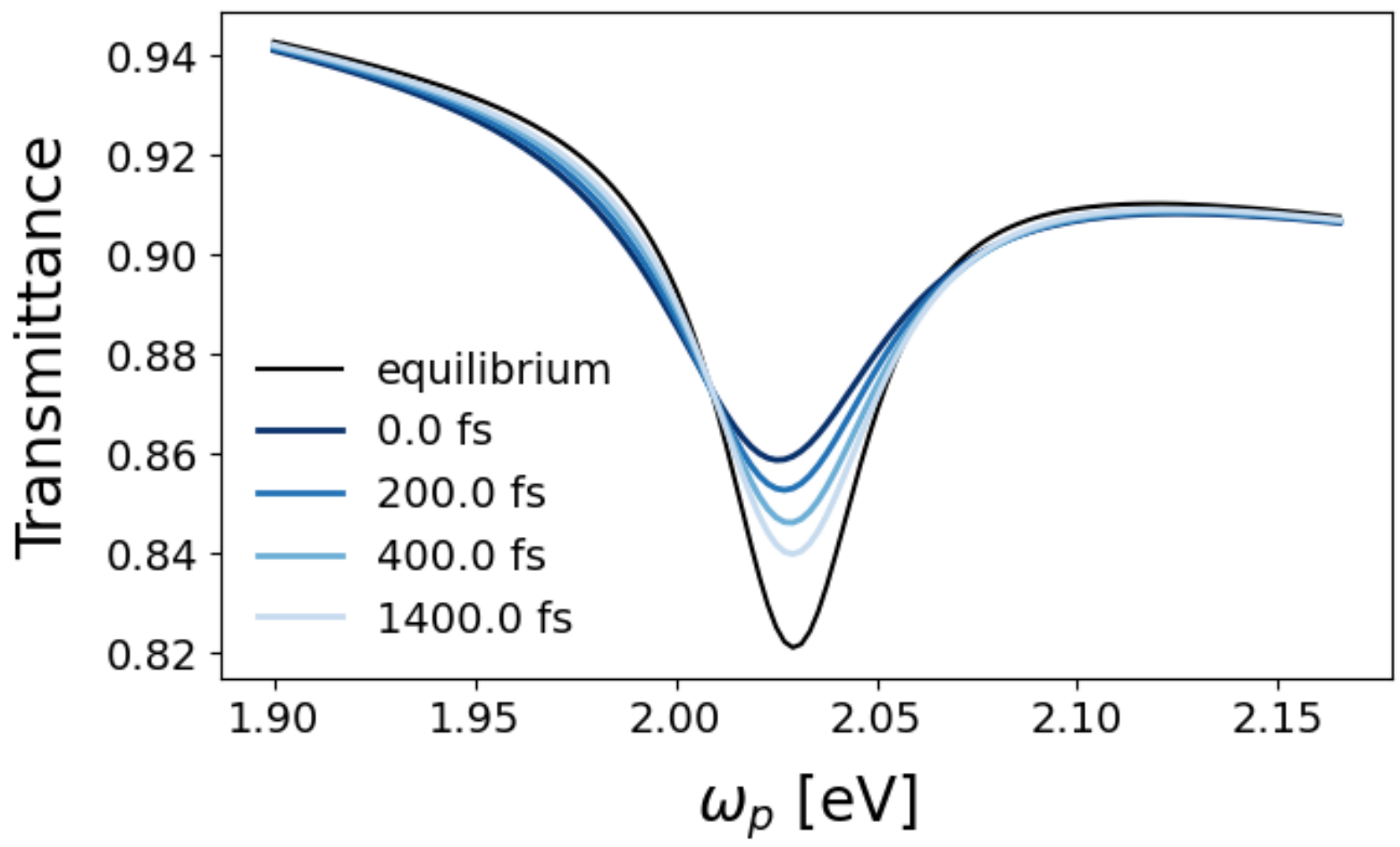

*Figure S12:* ***Time-dependent transmittance of the 1s exciton.*** *Calculated transmittance of 1L-$WS_2$ using the parameters obtained by the time-dependent fits. The black curve is the fitted equilibrium transmission (red line in Fig. S2b).*

# Supplementary Note 8: Long-time dynamics of the 1s exciton

We plot in Fig. S13 the transient $\Delta T/T$ of 1L-$WS_2$ measured upon photoexcitation by a pump pulse ($\omega_{pump} = 3.1$ eV) at fluence 20 $\mu J\ cm^{-2}$. The pump-induced changes to the transmissivity exhibit a long-lived response, reaching a plateau that remains visible at late time delays ($t = 25$ ps). This dynamics has been observed to persist over hundreds of picoseconds and is primarily dominated by interactions with the lattice, mediated by exciton-phonon scattering [34].

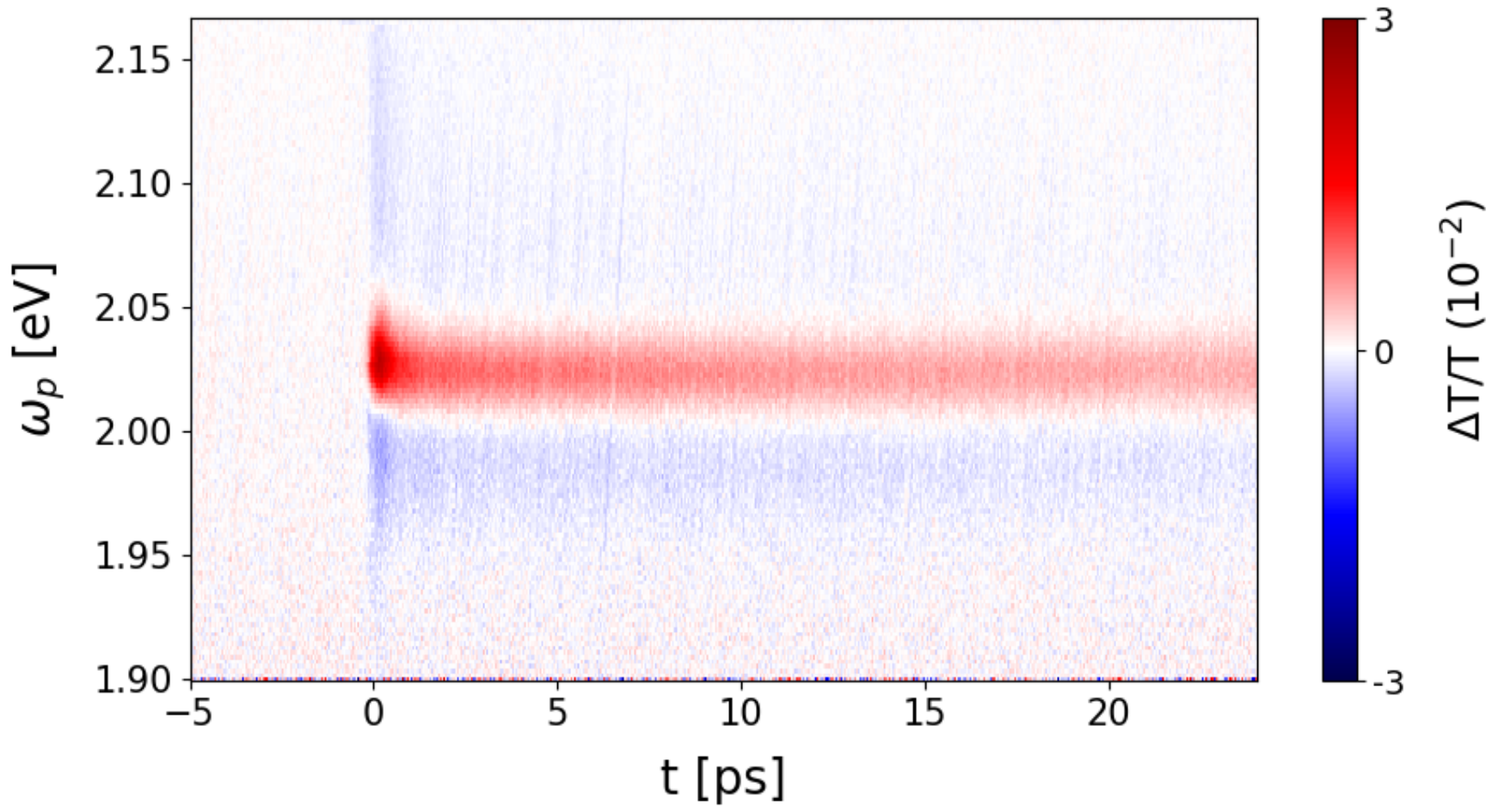

*Figure S13:* ***Long-time dynamics of the 1s exciton.*** *Time- and energy-resolved ΔT/T map upon photoexcitation by the visible pump at longer pump-probe delays.*

# Supplementary Note 9: Time-resolved Autler-Townes splitting

The measurements of the time-resolved Autler-Townes splitting have been performed according to the conceptual scheme in the rightmost panel of Fig. 1b. In order to simultaneously measure the pump- and the control field-induced changes in the transmittance, we used a differential acquisition based on two synchronized choppers. As detailed in Ref. [25], the visible pump and the mid-IR pulse have been mechanically chopped at frequencies 90 Hz and 45 Hz, respectively. The choppers are seeded by the same wavefunction generator so that their rotations are phase-locked. After the chopper, a small back-reflection of both pulses is routed to a pair of photodetectors to measure the presence/absence of the pump pulse. Since both photodetectors are digitized using the same PCI ADC digitizer (by Spectrum) that acquires pulse-by-pulse the probe and reference spectra, each probe/reference pulse can be assigned with a "label" which indicates the chopper status (ON/OFF). The choppers run one at twice the frequency of the other, so that the probe pulses can fall into four categories according to the status of each chopper: i) both pump pulses are on, ii) only the visible pump excites the sample, iii) only the mid-IR pump excites the sample, iv) no pump interacts with the sample.
To study the pump-induced dynamics, we delayed in time the mid-IR field with respect to the visible pump by the amount $\Delta t$. We then scanned the broadband probe in time with respect to the visible pump. For every $\Delta t$, the differential acquisition allowed us to compute three types of transient $\Delta T/T$ maps: one containing the response to both photoexcitations [(i-iv)/iv], one only measuring the changes induced by the visible pump [(ii-iv)/iv] and one measuring the Autler-Townes splitting induced only by the mid-IR field [(iii-iv)/iv].

In Fig. S14a we plot the transient $\Delta T/T$ maps of the first kind for selected time-delays $\Delta t$. For $\Delta t = -500$ fs (leftmost panel), the mid-IR field impinges on the sample 500 fs before the arrival of the visible pump. At $t = -500$ fs, the optical signature of the Autler-Townes splitting is clearly visible. From $t = 0$ fs on, the transmissivity is instead modulated by the interaction with the visible pump, according to the dynamics discussed Supplementary Note 7 (Fig. S10). At positive $\Delta t$, the mid-IR pulse interacts with the sample after the arrival of the above-gap pump.

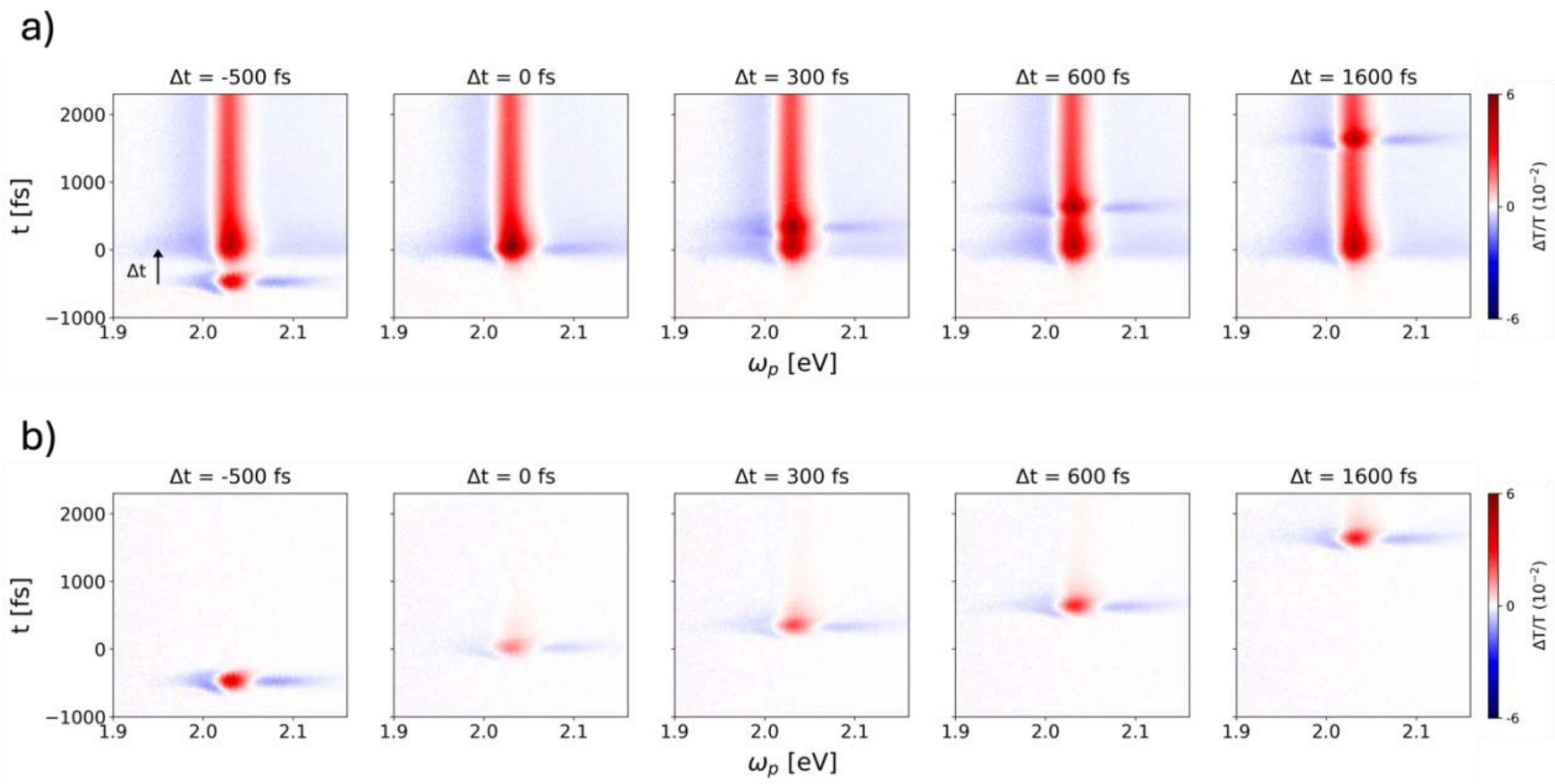

*Figure S14: **Three-pulse experiment in 1L-WS₂.** a) Time- and energy-resolved ΔT/T maps acquired upon simultaneous excitation by the above-gap pump ($\omega_{pump} = 3.1$ eV) and the mid-IR control field ($\omega_C = 155$ meV). The above-gap pump impinges at $t = 0$, while the mid-IR field is delayed in time by Δt with respect to it. b) Maps in a) after direct subtraction of the signal associated to the above-gap pump.*

In order to single out how the Autler-Townes splitting is affected by the excitation by the visible pump, we directly subtract from the maps in Fig. S14a the response associated to the visible pump only [(ii-iv)/iv]. The result of the subtraction is plotted in Fig. S14b. The map at $\Delta t = -500$ fs is obviously not affected by the subtraction because the mid-IR impinges on the sample before the visible pump. For $\Delta t \geq 0$, however, we observe a difference in the Autler-Townes splitting. Notably, we notice a quench of its intensity and a spectral broadening.
These effects are better visualized in the $\Delta T/T$ spectra plotted in Fig. S15. The colored lines are horizontal cuts of the maps in Fig. S14 at the corresponding time delay between the visible and the mid-IR pumps. To facilitate the comparison, we plot for each curve also the "reference" Autler-Townes spectrum (thin grey line), namely the spectrum acquired in the absence of the above-gap photoexcitation [(ii-iv)/iv]. The transparency window opened by the Autler-Townes effect is quenched by more than 50% at the temporal overlap between the visible pump and the mid-IR field. At larger delays, the intensity is then almost fully recovered.

We fitted the colored curves (for each time delay) using the complex susceptibility derived Eq. S3 to extract quantitative information on the changes induced by the above-gap pump. In particular, we used the following fitting function:

$$\frac{\Delta T}{T}(\Delta t) = \frac{T\left[\omega_p, \omega_{12}(\Delta t), \gamma_{12}(\Delta t)\right] - T\left[\omega_p, \omega_{12}(\Delta t), \gamma_{12}(\Delta t), \Omega_C = 0\right]}{T\left[\omega_p, \omega_{12}(\Delta t), \gamma_{12}(\Delta t), \Omega_C = 0\right]}$$

[Eq. S7]

Which has the same functional form of Eq. S6, but with the important difference that, at each time delay, the central frequency ($\omega_{12}$) and the width ($\gamma_{12}$) of the $1s$ transition are modified by the above-gap pump. We fed into the fitting function the temporal evolution $\omega_{12}(\Delta t)$ and $\gamma_{12}(\Delta t)$ extracted from the fit in Supplementary Note 7, namely the curves plotted in Fig. S10. The best fits are plotted as dashed black lines in Fig. S15.
For each time delay, the parameters providing the optimal fit to the data were extracted. The parameter $\gamma_{13}$, which was determined to be 65 meV under equilibrium conditions, exhibits only minor modifications under above-gap perturbation, increasing to 75 meV. However, no clear dependence of this parameter on the time delay $\Delta t$ was observed. In contrast, the parameter most significantly affected by photoexcitation is $\Delta_C$, indicating that the energy separation between the $1s$ and $2p$ states is dynamically modulated by the interaction with the visible pump. By determining the absolute redshift of the $1s$ energy level (Fig. S10b), we were able to calculate the corresponding absolute energy shift of the $2p$ level, as shown in Fig. 3a.

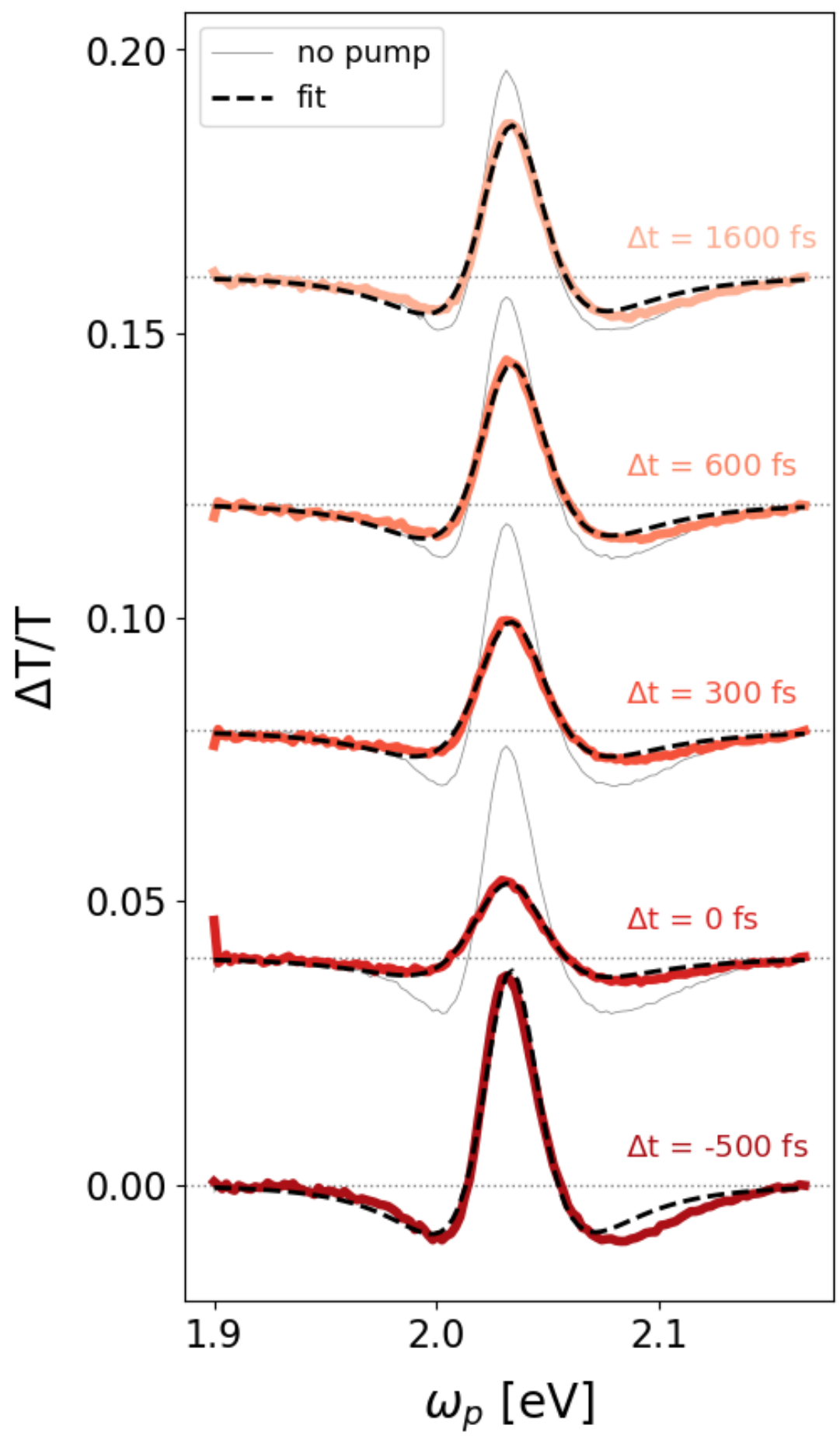

*Figure S15:* ***Time-dependent ATS spectra.*** *Spectra of the Autler-Townes splitting after the above-gap photoexcitation. Each colored curve corresponds to a different time-delay between the visible pump and the mid-IR control field that measures the splitting. For each delay, the thin grey curve is the splitting measured in the absence of the visible photoexcitation: it is the same for all the Δt and is a reference to enhance the pump-dependent changes. The dashed black line on top of each curve is the fit.*